\documentclass[manuscript]{aastex631}

\usepackage{graphicx}
\usepackage{tabularx}

\usepackage{tikz}
\usepackage{xurl}
\usepackage{xcolor}
\usepackage{ulem}

\renewcommand{\thetable}{\arabic{table}}

\begin{document}

\renewcommand{\thetable}{\arabic{table}}
\setcounter{table}{0}

\title{Impacts of Atmospheric Carbon Species and Stellar Type on Climates of Terrestrial Planets}

\shorttitle{Impacts of Carbon Species on Climate}{Impacts of Carbon Species on Climate}

\author[0009-0002-1943-4709]{Jared Landry}
\affiliation{Earth-Life Science Institute, Institute of Science Tokyo, 2-12-1 Ookayama, Meguro, Tokyo, 152-8550, Japan}
\email{landryj@elsi.jp}

\author[0000-0003-1965-1586]{Hiroyuki Kurokawa}
\affiliation{Department of General Systems Studies, The University of Tokyo, 3-8-1 Komaba, Meguro, Tokyo 153-8902, Japan}
\affiliation{Department of Earth and Planetary Science, The University of Tokyo, Tokyo 113-0033, Japan}

\author[0000-0002-6602-7113]{Tetsuo Taki}
\affiliation{Department of General Systems Studies, The University of Tokyo, 3-8-1 Komaba, Meguro, Tokyo 153-8902, Japan}

\author[0000-0002-2786-0786]{Yuka Fujii}
\affiliation{Division of Science, National Astronomical Observatory of Japan, 2-21-1 Osawa, Mitaka, Tokyo 181-8588, Japan}
\affiliation{Graduate Institute for Advanced Studies, SOKENDAI, 2-21-1 Osawa, Mitaka, Tokyo 181-8588, Japan}

\author{Kosuke Aoki}
\affiliation{Department of Earth and Planetary Sciences, Institute of Science Tokyo, Tokyo 152-8551, Japan}

\author[0000-0001-6702-0872]{Hidenori Genda}
\affiliation{Earth-Life Science Institute, Institute of Science Tokyo, 2-12-1 Ookayama, Meguro, Tokyo, 152-8550, Japan}



\begin{abstract}

The climates of terrestrial planets are largely determined by the composition of their atmospheres and spectral types of their host stars. Previous studies suggest a wide range of carbon species abundances (CO\textsubscript{2}, CO, and CH\textsubscript{4}) can result from variations in reducing fluxes and stellar spectral types which influence photochemistry. However, a systematic investigation of how varying carbon species, particularly CO, affect planetary climates across wide parameter spaces remains limited. Here, we employ a one-dimensional radiative-convective equilibrium model to examine the dependence of planetary climate on the abundances of carbon species and host star type. We find that CO, due to weak absorption of stellar radiation, induces only moderate changes in stratospheric temperature, while its effect on surface temperature is negligible. Under Earth-like $p_\mathrm{N_2}$ (where $p_\mathrm{i}$ is the partial pressure on the surface of species i), for cases with fixed $p_\mathrm{CO_2}$, increase in CO leads to surface cooling on planets orbiting Sun-like stars unless the sum of $p_\mathrm{CO_2}$ and $p_\mathrm{CH_4}$ exceeds $\sim$1 bar. Whereas it results in surface warming for planets around M-type stars. When the total pressure of carbon species is fixed, converting CO\textsubscript{2} or CH\textsubscript{4} into CO always induces cooling. These effects arise from a combination of CO Rayleigh scattering, pressure broadening of greenhouse gas absorption lines, and varying water vapor levels. We further discuss how CO- and CH\textsubscript{4}-driven cooling (warming) can trigger positive (negative) climate-photochemistry feedback, influencing atmospheric evolution. Additionally, we suggest CO-rich planets may be less susceptible to water loss and atmospheric oxidation due to lower stratospheric water vapor content. %250 words < 250 words limit

\end{abstract}



\section{Introduction} \label{sec:intro}

The composition of an atmosphere controls a planet’s climate and thus it's habitability because the various gases behave differently via scattering and absorbing stellar and planetary radiation \citep[e.g.,][]{kasting1986climatic,haqq2008revised,catling2017atmospheric}. The solar-system terrestrial planets (Venus, Earth, and Mars) today and in the past exhibit a variety of atmospheres. Due to variations in host star and planetary properties, extrasolar terrestrial planets can have even greater diversity. Here we focus on carbon species (CO$_2$, CO, and CH$_4$) as they would typically be major atmospheric constituents.

The abundance of these carbon species is controlled by planetary processes. The CO$_2$ abundance in Earth's atmosphere is thought to have been controlled by the carbonate-silicate cycling: a balance between volcanic supply and loss through silicate weathering, carbonate precipitation, and subduction \citep{walker1981negative, foley2015role, catling2017atmospheric}. Thanks to the temperature dependence of the weathering rate, the carbonate-silicate cycling acts as a thermostat \citep{walker1981negative}. This negative feedback has likely contributed to keep early Earth habitable \citep{walker1981negative,Krissansen-Totton+2018PNAS..115.4105}, and is considered to determine the extent of habitable zones \citep{Kasting+1993Icar..101..108K,Kopparapu+2013ApJ...765..131K}.

Carbon speciation including CO and CH$_4$ is determined by the supply and loss of reducing power via volcanic outgassing and hydrogen escape, as well as atmospheric photochemistry. An oxidizing mantle such as modern Earth's releases CO$_2$, while reducing systems release CO and CH$_4$ \citep{catling2017atmospheric}. It has been proposed that Earth's mantle was more reducing in the Archean, releasing reduced gases \citep{Aulbach+2016,nicklas2019secular,Kadoya+2020}, although the interpretation of geochemical data is controversial \citep{Zhang+2024}. For rocky bodies in the solar-system, oxygen fugacity spans a range of $\sim 10$ orders of magnitude \citep{Cartier+Wood2019}. 

Reducing power is lost from a planet via hydrogen escape to space. While various escape mechanisms dictate the hydrogen escape rate,  on Earth and elsewhere in the solar-system, the escape of hydrogen is thought to be limited by the rate at which it can diffuse to the upper atmosphere, where it then escapes \citep{catling2017atmospheric}. This is called diffusion-limited escape. On modern Earth, the supply of hydrogen to the upper atmosphere depends largely on the amount of water vapor in the stratosphere \citep{Hunten1973JAtS...30.1481H,catling2017atmospheric}. When the mixing ratio of water vapor in the stratosphere increases to a certain threshold, the escape of hydrogen is no longer limited by its supply to the upper atmosphere and is instead limited by the energy supply flux from X-ray and extreme ultraviolet (XUV) radiation, called energy-limited escape \citep{Watson+1981Icar...48..150W,catling2017atmospheric}.
Thus, the climate affects the rate of hydrogen escape through determining the water vapor content in the stratosphere.

Finally, photochemistry induced by ultraviolet (UV) irradiation from the host star also controls carbon speciation. While CO\textsubscript{2} dissociates to CO and O with $\lesssim$200 nm radiation, their direct recombination is spin-forbidden \citep{catling2017atmospheric}. A catalytic cycle with OH radicals, formed with dissociation of H$_2$O, stabilizes CO$_2$ in habitable worlds with CO$_2$-rich atmospheres \citep{McElroy+Donahue1972Sci...177..986M,catling2017atmospheric}. However, the runaway dissociation of CO$_2$ (also called as the CO runaway) may be triggered on planets orbiting M-type stars with less H$_2$O-dissociating photons \citep{Tian+2014,harman2015abiotic} and with colder climates such as early Mars where the water vapor content is lower \citep{zahnle2008photochemical}. \cite{watanabe2024relative} recently performed extensive parameter survey for the volcanic reducing flux and stellar type and summarized the conditions for CO runway. 

In climate modeling, CO\textsubscript{2}, as it is the most abundant greenhouse gas in the current Earth’s atmosphere aside from water vapor, has been the focus of most climate studies that contain carbon compounds \citep{Manabe+Wetherald1967,kasting1986climatic,wordsworth2013water,ramirez2014can}. Since CO\textsubscript{2} has strong absorption bands in the wavelength range for planetary black-body radiation, increasing CO\textsubscript{2} levels typically increases surface temperatures 
\citep{kasting1986climatic,wordsworth2013water,ramirez2014can}. However, there are also situations where increasing CO\textsubscript{2} content may decrease temperatures. Here, water content decreases with increasing $p_\mathrm{CO_2}$ (hereafter $p_i$ represents the partial pressure of the species $i$ at the surface) because the sensible heat (energy required for a change in temperature) of the atmosphere exceeds the latent heat (energy required for a change of phase). When the sensible heat of an expanding, upwelling air parcel is greater than latent heat, the work being done is supplied by internal energy rather than by the release of latent heat from a change of phase. This leads to a decrease in temperature, and thus limits the amount of water vapor in the upper atmosphere \citep{wordsworth2013water}. In addition to this water vapor effect, the CO\textsubscript{2} greenhouse effect eventually saturates at high levels while Rayleigh scattering does not, leading to a decrease in temperatures \citep{kasting1993earth}. This phenomenon is responsible for setting an outer edge to the habitable zone of stellar systems, otherwise CO\textsubscript{2} could be added continuously to counteract the diminishing top-of-atmosphere (TOA) flux with increasing orbital radii.

Studies investigating methane’s influence on climate have also been performed \citep{pavlov2000greenhouse,haqq2008revised,arney2016pale, arney2017pale}. Methane has direct and indirect effects on the atmosphere; absorption as a greenhouse gas \citep{pavlov2000greenhouse,Ramirez+2018}, and scattering from haze produced by CH\textsubscript{4} via photochemical reactions \citep{haqq2008revised,arney2016pale}. It was found that increasing the CH$_4$ content causes warming up to $f_\mathrm{CH_4}/f_\mathrm{CO_2} \lesssim 0.1$
\citep[hereafter $f_i$ represents the volume mixing ratio of the species $i$,][]{pavlov2000greenhouse,haqq2008revised}. However, when $f_\mathrm{CH_4}/f_\mathrm{CO_2} \gtrsim 0.1$, photochemical hazes form in the atmosphere as a result of CH\textsubscript{4} photochemical reactions, and these hazes can lead to significant cooling due to scattering of incoming radiation for planets orbiting G-type stars, or possible warming from absorption of outgoing radiation for planets orbiting M-type stars
\citep{arney2017pale}.The inclusion of hazes, however, is beyond the scope of this study.

Compared to CO$_2$ and CH$_4$, the climate effects of CO have received less attention. However, photochemical modeling shows that CO can be a major carbon species on planets orbiting M-type stars \citep{Tian+2014,harman2015abiotic,Hu+2020} and on early Mars \citep{zahnle2008photochemical}, the latter of which is also supported from carbon-isotopic evidence \citep{Ueno+2024}. Because CO is infrared-inactive, the primary influences of CO are thought to be scattering of stellar irradiation, which causes a decrease in temperatures. On the other hand, CO causes pressure-broadening of absorption lines of co-existing greenhouse gases, which increases temperatures. \cite{Hu+2020} modeled climates of CO-rich atmospheres by using another non-greenhouse gas, N$_2$, instead of CO. However, CO has weak but non-negligible absorption lines in near- to mid-infrared wavelengths \citep{gordon2017hitran2016}, whose effects on the atmospheric temperature profile remains unexplored. Moreover, a systematic investigation of how varying carbon species, particularly CO, affect planetary climates across a wide parameter space remains limited. In the extensive survey for the CO runaway by \cite{watanabe2024relative}, a fixed atmospheric temperature profile is used. Therefore, the climate effects of the CO production and their feedback on photochemistry remains to be investigated.

We aim to perform a comprehensive study on the climate effects of carbon species including CO as well as CO\textsubscript{2} and CH\textsubscript{4}, and discuss the feedback on atmospheric chemistry. Here, we utilize a one-dimensional (1D) radiative-convective equilibrium model to understand the relationships between the planetary climate and atmospheric carbon species (CO\textsubscript{2}, CO, CH\textsubscript{4}) , as well as the spectral type of the host star. Furthermore, we discuss how these different dependencies of the climate influence photochemical feedback on the atmospheric oxidation state, to understand how atmospheres evolve over time. We also discuss the fate of surface water based on the calculated stratospheric water content and associated hydrogen escape to space.

\section{Methods} \label{sec:Methods}

\subsection{Temperature and water vapor profiles} \label{subsec:methods:profiles}

To calculate the 1D temperature and water vapor profiles of the atmosphere for given sets of partial pressures of carbon species and spectral types of host stars, we used CLIMA, a module of the ATMOS code\footnote{\url{https://github.com/VirtualPlanetaryLaboratory/atmos}} \citep{kasting1984response,arney2016pale}. Because the original CLIMA only considers CO\textsubscript{2} and CH\textsubscript{4} for carbon species, we newly included the effect of CO on absorption and scattering of radiation, and on the heat capacity.

In the radiative-convective equilibrium model adapted by CLIMA, the atmosphere consists of two layers: the stratosphere and the troposphere. The troposphere is characterized by convection and its temperature and water vapor profiles are modeled with the moist adiabat. Convection is assumed for layers where the temperature lapse rate ($-dT/dz$, where $T$ is the temperature and $z$ is the height from the surface) is larger than that of the adiabat. This procedure finds the tropopause, the boundary between the troposphere and stratosphere. The moist adiabatic lapse rate takes the release of the latent heat from condensation of saturated water vapor into account \citep{Ingersoll1969,kasting1984response,kasting1988runaway,catling2017atmospheric}. Given the saturation vapor pressure of each layer calculated from the adiabat, the mixing ratio of water vapor is then obtained by adapting a relative humidity (RH) model. The original CLIMA code assumes the RH model of \cite{manabe1967thermal}, which is developed to mimic that of current Earth. Since we consider planets with higher temperature where higher RH is expected \citep{Goldblatt+2013NatGe...6..661G,ramirez2014can}, we adopt the model of \cite{kasting1986climatic}. To add CO in the convection model, its heat capacity is taken from the NIST database \footnote{\url{https://webbook.nist.gov/cgi/cbook.cgiID=C630080&Type=JANAFG&Table=on}}.

We utilize two different models; the \textquotedblleft forward\textquotedblright\ and \textquotedblleft inverse\textquotedblright\ models. In the forward model, the temperature profile in the stratosphere is modeled with a profile which satisfies the net radiation flux (the sum of incoming and outgoing flux) equal to zero. Here, the tropopause is determined via a minimum in the temperature profile we calculate.

In the inverse model, the stratospheric temperature profile is assumed to be isothermal. Here the water vapor mixing ratio in the stratosphere is assumed to be constant because of the temperature dependence of the saturation vapor pressure. Here, the tropopause is found where the adiabatic temperature profile reaches an assumed temperature of the stratosphere. Following previous studies \cite[e.g.,][]{ramirez2014can}, we assumed 200 K, which roughly corresponds to Earth's skin temperature \citep[][]{catling2017atmospheric}.

As we show in Section \ref{subsec:results:COabs}, the results of forward and inverse calculations show only a moderate difference in the stratosphere, while the surface temperature is not affected significantly. Moreover, forward calculations are computationally more expensive and sometimes cause difficulty in finding equilibrium solutions. Thus, we performed forward modeling for limited parameter sets to investigate the effect of CO absorption on stratospheric temperature and water vapor content and to validate the inverse modeling (see Section \ref{subsec:results:COabs}). Then we performed an extensive parameter survey with inverse modeling (Sections \ref{subsec:results:COCH4}, \ref{subsec:results:CO2}, \ref{subsec:results:Spec}, and \ref{subsec:results:pCtotal}). 

\subsection{Radiative transfer} \label{subsec:methods:RT}

\begin{figure}[ht]
    \centering
    \includegraphics[width=0.75\textwidth]
    {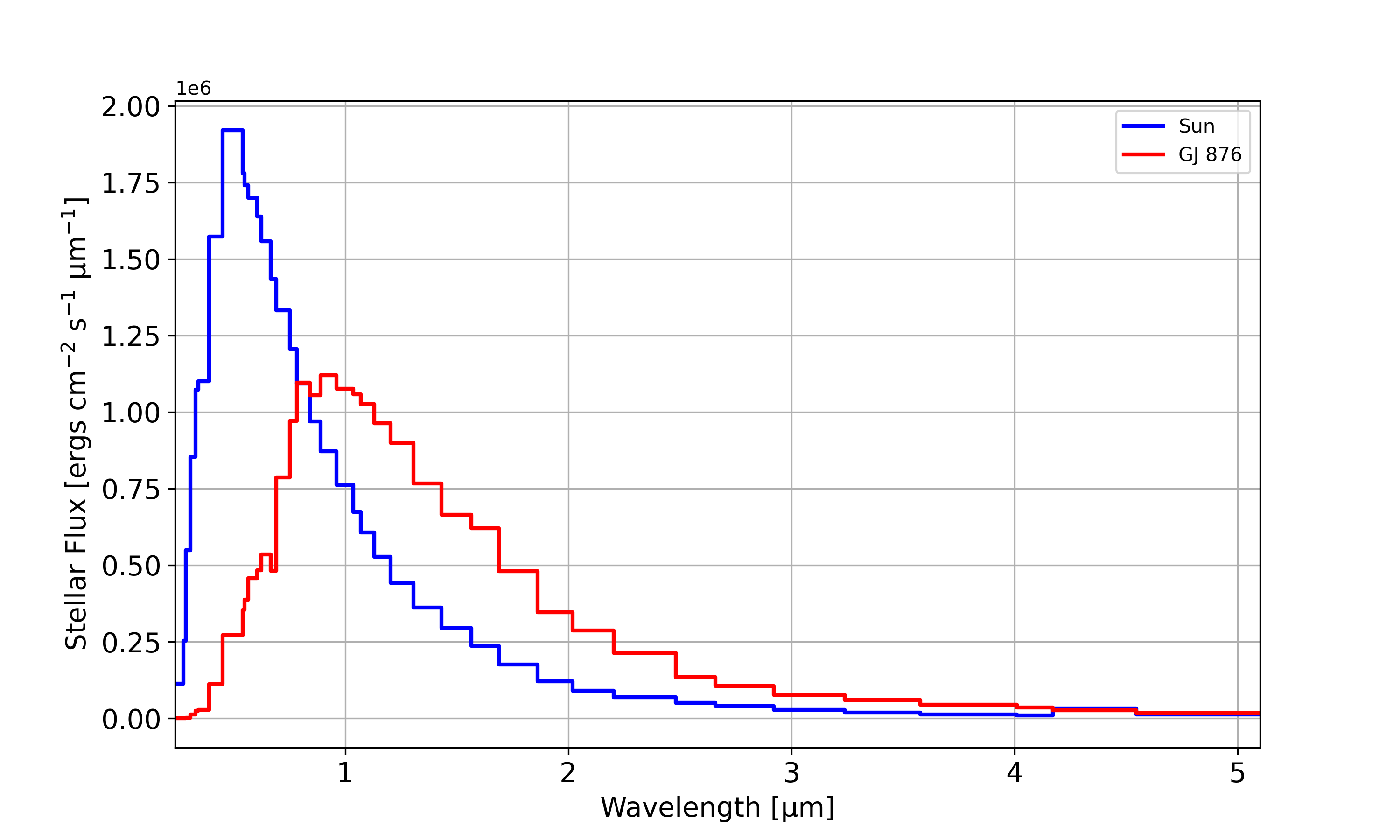}
    \caption{Spectra of the Sun and GJ 876 for the wavelength range $0.25$--$5.4$ $\mu$m used in our calculations, divided by the wavelength range in each bin, and found at the top of the atmosphere. The Sun case is found for an Earth-like terrestrial planet orbiting at 1 AU, and the GJ 876 case is found for an Earth-like terrestrial planet at an orbital distance where 0.003 bar CO\textsubscript{2} and 0.8 bar N\textsubscript{2} in the atmosphere yields an equilibrium surface temperature of 285 K, consistent with modern Earth.}
    \label{fig:methods:spectra}
\end{figure}

Radiative transfer in CLIMA is broken into two different bandpasses: stellar and planetary (thermal) radiation \citep{kasting1984response,arney2016pale}. 
The transmissivity for each individual gas is determined by the correlated-$k$ method \citep{kato1999k}, where absorption of photons in different wavelengths within a single bin is only treated statistically to maximize computational efficiency while maintaining accuracy in the flux for the wavelength bin. The two-stream approximation is assumed for the calculation of radiative fluxes into and out of each layer of the atmosphere\citep{toon1989rapid}.

Because the version of CLIMA we are based on originally only contains CO\textsubscript{2}, CH\textsubscript{4}, and H\textsubscript{2}O as absorbing gases, we added CO to the code for this study. CO is a weak absorber, but the strongest absorption feature of CO occurs outside of the stellar wavelength range considered in the original CLIMA \citep{kasting1984response,rothman2009hitran,catling2017atmospheric}.
For the stellar wavelength range, originally CLIMA ranges from 0.25 to 4.5 ${\rm \mu m}$, broken up into 38 wavelength bins 
\citep{kasting1984response,arney2016pale}, but because of CO’s absorption spectrum, we extended the range to 5.4 ${\rm \mu m}$ and added an additional wavelength bin, making 39 total. We calculated the coefficients of the correlated-$k$ method for spectral bins including the new bin. The absorption data for the original gases in CLIMA comes from HITRAN 2008, except for H\textsubscript{2}O, which comes from HITEMP 2010, and the data for CO is taken from the HITRAN 2016 database \citep{rothman2009hitran,gordon2017hitran2016}. The thermal-infrared spectrum is broken up into 55 bins from 0.7 to 500 ${\rm \mu m}$ \citep{kasting1984response,arney2016pale}. Finally, the scattering parameters for CO are taken from \cite{vardavas1984solar}. 

\subsection{Input parameters and calculation procedures} \label{subsec:methods:input}

The inputs into the CLIMA code include the partial pressures of the non-condensible gases at the surface, the radiation spectrum of the host star, the incoming stellar flux, the Bond albedo of the surface $A$, the gravitational acceleration $g$, and the pressure at the top of the atmosphere $p^\mathrm{TOA}$. 
The partial pressures of carbon species, host stellar type and incoming flux were varied as detailed below. For the other input parameters, we assumed fixed values. We assumed $p_\mathrm{N_2} = 0.8\ \mathrm{bar}$, $A = 0.32$ which is consistent for the assumption of modern Earth modeled without clouds, $g = 9.8\ \mathrm{m\ s^{-2}}$, and $p^\mathrm{TOA} = 10^{-6}\ \mathrm{bar}$.

Because the mixing of different gases changes their partial pressures from those obtained when they exist alone, we define two partial pressures: the partial pressure in the mixed gas $\tilde{p}_i$ and the pressure when it exists by itself $p_i$, where the subscript $i$ represents a gas species of interest,  consistent with previous studies \citep{way+2017} (The latter is our input parameter and is hereafter what we are concerned with when discussing the partial pressure of carbon species. The prior is only used within our code for calculations). This is necessary because the atmosphere is a mixture of gasses with varying molar weights, and $\tilde{p}_i$ and $p_i$ are only equal when each individual gas has the same molar weight. Those two partial pressures can be converted as follows. The partial pressure in the mixed gas $\tilde{p}_i$ is defined as,
\begin{equation}
    {\tilde{p}_i} = {f_i} {p},
    \label{eq:1}
\end{equation}
where $p$ is total pressure, which is given by,
\begin{equation}
    p = p_\mathrm{H_2O} + p_\mathrm{CO_2} + p_\mathrm{CO} + p_\mathrm{CH_4} + p_\mathrm{N_2}.
\end{equation}
The mixing ratio $f_i$ in Equation \ref{eq:1} is described by,
\begin{equation}
    f_i = \frac{\sigma_i}{\sigma_\mathrm{H_2O} + \sigma_\mathrm{CO_2} + \sigma_\mathrm{CO} + \sigma_\mathrm{CH_4} + \sigma_\mathrm{N_2}},
\end{equation}
where $\sigma_i$ is the column density, given by,
\begin{equation}
    {\sigma_i} = \frac{p_i}{m_i g}.
\end{equation}
Here $m_i$ is the molecular mass of species $i$.

\setlength{\abovecaptionskip}{15pt}
\setlength{\belowcaptionskip}{0pt}
\setlength{\tabcolsep}{30pt}
\setcounter{table}{0}
\begin{table}[h]
    \centering
    \resizebox{\textwidth}{!}{
        \begin{tabular}{ccccc}
            \hline
            \hline
            & \textbf{Fixed $p_\mathrm{CO_2}$ case} & & \textbf{Fixed \textit{p}\textsubscript{Ctotal} case} & \\
            \hline
            \textbf{Variable} & \textbf{Range} & & \textbf{Variable} & \textbf{Range} \\
            \hline
            $p_\mathrm{CO}$ [bar] & $10^{-6}$--$10^{2}$ & & $p_\mathrm{CO}/p_\mathrm{CO_2}$ & $10^{-6}$--$10^{2}$ \\
            \hline   
            $p_\mathrm{CH_4}$ [bar] & $10^{-6}$--$10^{2}$ & & $p_\mathrm{CH_4}/p_\mathrm{CO_2}$ & $10^{-6}$--$10^{2}$ \\
            \hline   
            $p_\mathrm{CO_2}$ [bar] & $10^{-2}, 1$ & & \textit{p}\textsubscript{Ctotal} [bar] & 0.1 \\
            \hline   
            Host Star & Sun, GJ 876 & & Host Star & Sun \\
            \hline
        \end{tabular}
    }
    \caption{Input parameters for our study.}
    \label{tab:parameters}
\end{table}

For the partial pressures of the three carbon species, we made our calculations for two cases to summarize our results in two-dimensional parameter spaces (Table \ref{tab:parameters}). The first is where we fixed $p_\mathrm{CO_2}$ and treated $p_\mathrm{CO}$ and $p_\mathrm{CH_4}$ as independent parameters. This case mimics a situation where $p_\mathrm{CO_2}$ is controlled by the carbonate-silicate cycling \citep{walker1981negative,foley2015role}, while $p_\mathrm{CO}$ and $p_\mathrm{CH_4}$ vary due to changing the reducing flux. We note that this is simplification; changing $p_\mathrm{CO}$ and $p_\mathrm{CH_4}$ can change the surface temperature, which then influences the weathering rate and thus $p_\mathrm{CO_2}$ in reality.

The second case is a closed system where the total pressure of carbon species ($p_\mathrm{Ctotal}\equiv p_\mathrm{CO_2}+p_\mathrm{CO}+p_\mathrm{CH_4}$) is fixed, mimicking a system without an active carbonate-silicate cycle. We expect such cases to be possible because the carbonate-silicate cycle is largely dependent on tectonic activity. If a terrestrial planet is tectonically inactive, there can be limited burial or degassing of carbon species. In addition to Earth and perhaps other solar system terrestrial planets earlier in their history, such worlds likely exist beyond our solar system, giving us cause to include such situations into our model.However, it is worth noting that studies have investigated the ability for terrestrial planets with stagnant lids to maintain a carbon cycle \citep[e.g.,][]{foley+2018}. 

We consider two different host stars: the Sun and a M-type star GJ 876, whose spectra can be found in Figure \ref{fig:methods:spectra}. For the former case, we assumed that a planet receives an incoming flux equivalent to that of current Earth. For the latter case, we assumed an incoming flux with which a planet with $p_\mathrm{N_2} = 0.8\ \mathrm{bar}$ and $p_\mathrm{CO_2} = 0.003\ \mathrm{bar}$ has an equilibrium surface temperature of $285\ \mathrm{K}$ (consistent with modern Earth). At the top of the atmosphere in this case, the incoming flux is 1192.5 W/{m$^2$}, equal to 0.8726 times the solar constant.

For each parameter set, we determined the atmospheric structure in equilibrium by changing the surface temperature iteratively. 
We determined that an equilibrium solution was obtained when the relative difference between incoming and outgoing flux at the top of the atmosphere fell below $10^{-3}$. We note that, in cases with very high $p_\mathrm{CH_4}$ ($\gtrsim 10\ \mathrm{bar}$) and the M-type host star, the equilibrium surface temperature exceeds 646.96 K, the triple point of H\textsubscript{2}O. Because we are only interested in potentially habitable worlds (i.e., worlds with liquid water on the surface), we excluded those cases.

\section{Results} \label{sec:results}

\subsection{The effect of CO absorption on temperature and water vapor profiles}
\label{subsec:results:COabs}

\setlength{\abovecaptionskip}{15pt}
\setlength{\belowcaptionskip}{0pt}

\begin{figure}[ht]
    \centering
    % First image (a)
    \begin{minipage}[b]{0.49\textwidth}
        \centering
        \includegraphics[width=\textwidth]{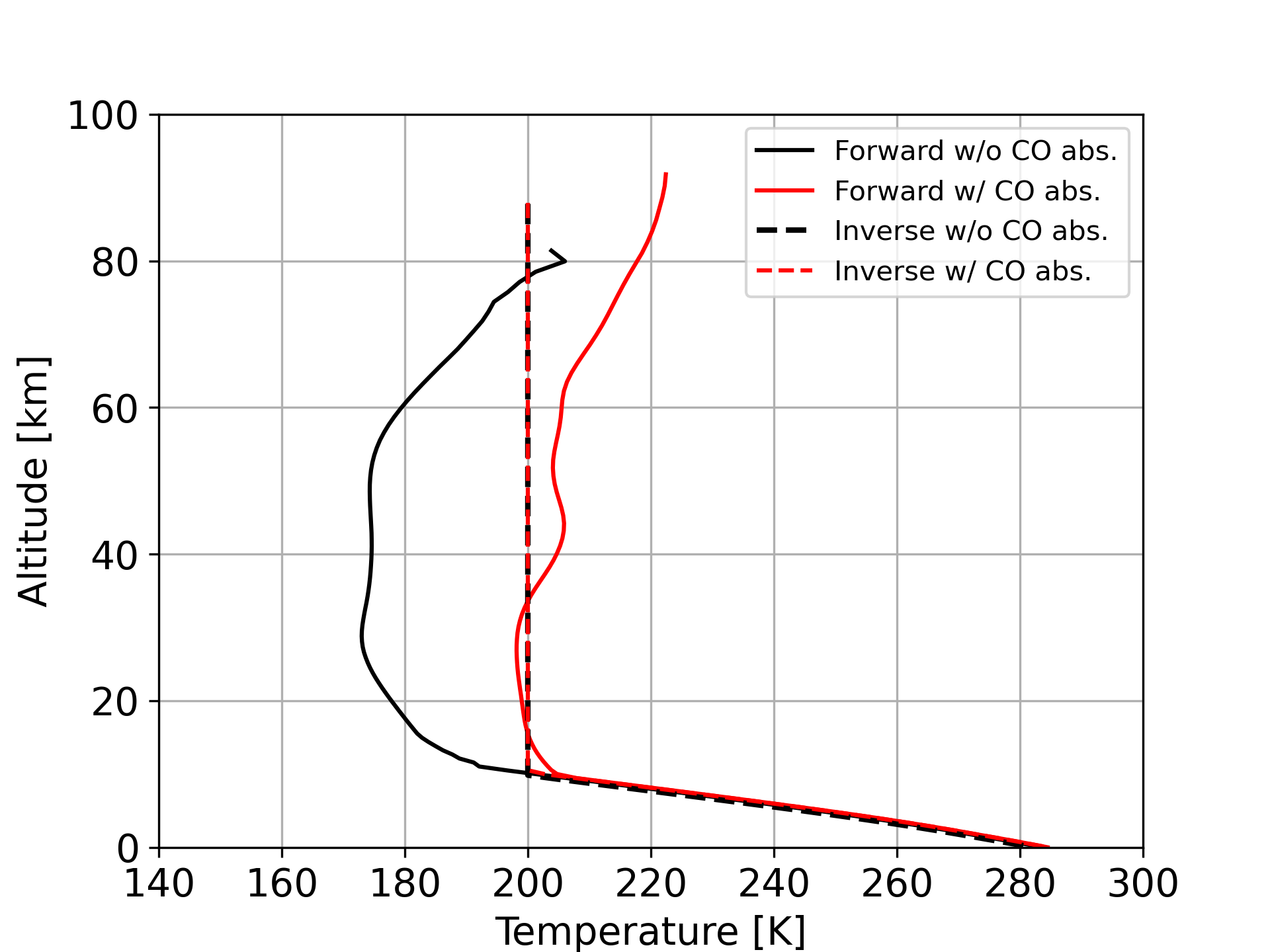}
        \put(-235,180){\scriptsize \textbf{(a)}}
    \end{minipage}
    \hfill
    % Second image (b)
    \begin{minipage}[b]{0.49\textwidth}
        \centering
        \includegraphics[width=\textwidth]{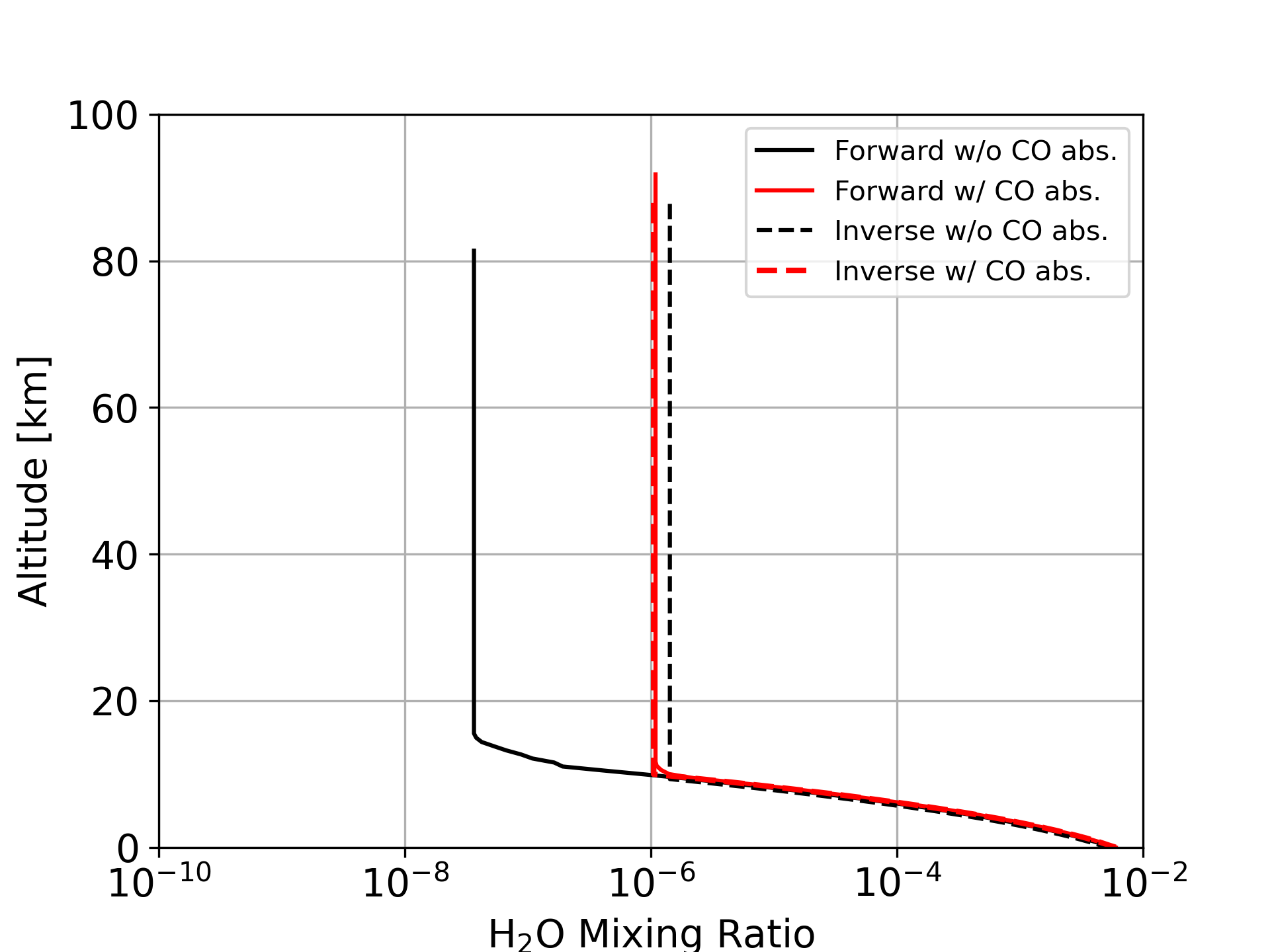}
        \put(-235,180){\scriptsize \textbf{(b)}}
    \end{minipage}

    \caption{\footnotesize Temperature (a) and water vapor (b) profiles for $p_\mathrm{CO} = 1\ \mathrm{bar}$ and $p_\mathrm{CO_2} = 10^{-3}\ \mathrm{bar}$ with the Sun being the host star. Solid red, solid black, dashed red, and dashed black lines show the results obtained with the forward model with CO absorption, forward model without CO absorption, inverse model with CO absorption, and inverse model without CO absorption, respectively.}
    \label{fig:results:profiles}
\end{figure}

The predominant effect of CO absorption occurs as moderate heating of the stratosphere. Figure \ref{fig:results:profiles} shows the temperature and water vapor profiles for a CO-rich case ($p_\mathrm{CO} = 1\ \mathrm{bar}$ and $p_\mathrm{CO_2} = 10^{-3}\ \mathrm{bar}$ with the Sun as the host star) obtained with the forward and inverse modeling, where a result without CO absorption is shown for comparison. We found that, without CO absorption, the stratospheric temperature is lower by 20--30 K in the forward model than Earth's skin temperature assumed in the inverse model (Figure \ref{fig:results:profiles}a). This is caused by the non-gray effect of the atmosphere \citep{Pierrehumbert2010ppc..book.....P,wordsworth2013water,Goessling+Bathiany2016ESD.....7..697G}. In a gray atmosphere, the temperature at the top of the atmosphere follows the skin temperature. In contrast, in a non-gray atmosphere with an atmospheric window, radiation in the window wavelengths travels directly from the surface or troposphere to space through the stratosphere. To maintain radiative balance, the upward radiative flux in the opaque bands must weaken, which results in persistent stratospheric cooling. In the forward model with CO absorption, however, the stratospheric temperature recovers. This is because of absorption of incoming stellar radiation by CO and associated heating, while the similarity in the stratospheric temperature with that in the inverse model is a coincidence. As a result, the stratospheric water-vapor content is comparable between the forward model with CO absorption and the inverse model, while it is lower in the forward model without CO absorption (Figure \ref{fig:results:profiles}b). Because of the exponential dependence of saturation vapor pressure on the temperature, the modest change in the cold trap temperature impacts the H\textsubscript{2}O mixing ratio by an order of magnitude. We note that the cold trap is defined as where the saturation mixing ratio of water vapor reaches a minimum due to the temperature decrease with altitude, limiting the transport of water vapor to the upper atmosphere and forcing remixing with the lower atmosphere \citep{catling2017atmospheric}. In our model we set the boundary between the convective troposphere and the isothermal radiative stratosphere (i.e. the tropopause) as the cold trap.

In contrast, the surface temperature is unchanged between the three cases (Figure \ref{fig:results:profiles}a). This allows us to utilize the inverse calculations for our parameter survey. Our interests lie largely on surface conditions, as they are most relevant for a planet's habitability. However, we note that this simplification may have some impacts on the stratospheric water content and the discussion on water loss. We discuss the influences of applying the inverse model on our results in Section \ref{subsec:discussion:limitation}. In the following sections, we show the dependence of surface temperature and water vapor content on partial pressures of carbon species and on the stellar type, obtained with the inverse model.

\subsection{\texorpdfstring{Dependence on CO and $CH_\mathrm{4}$ Partial Pressure}{Dependence on CO and CH4 Partial Pressure}}\label{subsec:results:COCH4}

\setlength{\abovecaptionskip}{0pt}
\setlength{\belowcaptionskip}{0pt}

\begin{figure}[ht]
    \centering
    \begin{minipage}[b]{0.49\textwidth}
        \centering
        \includegraphics[width=\textwidth]{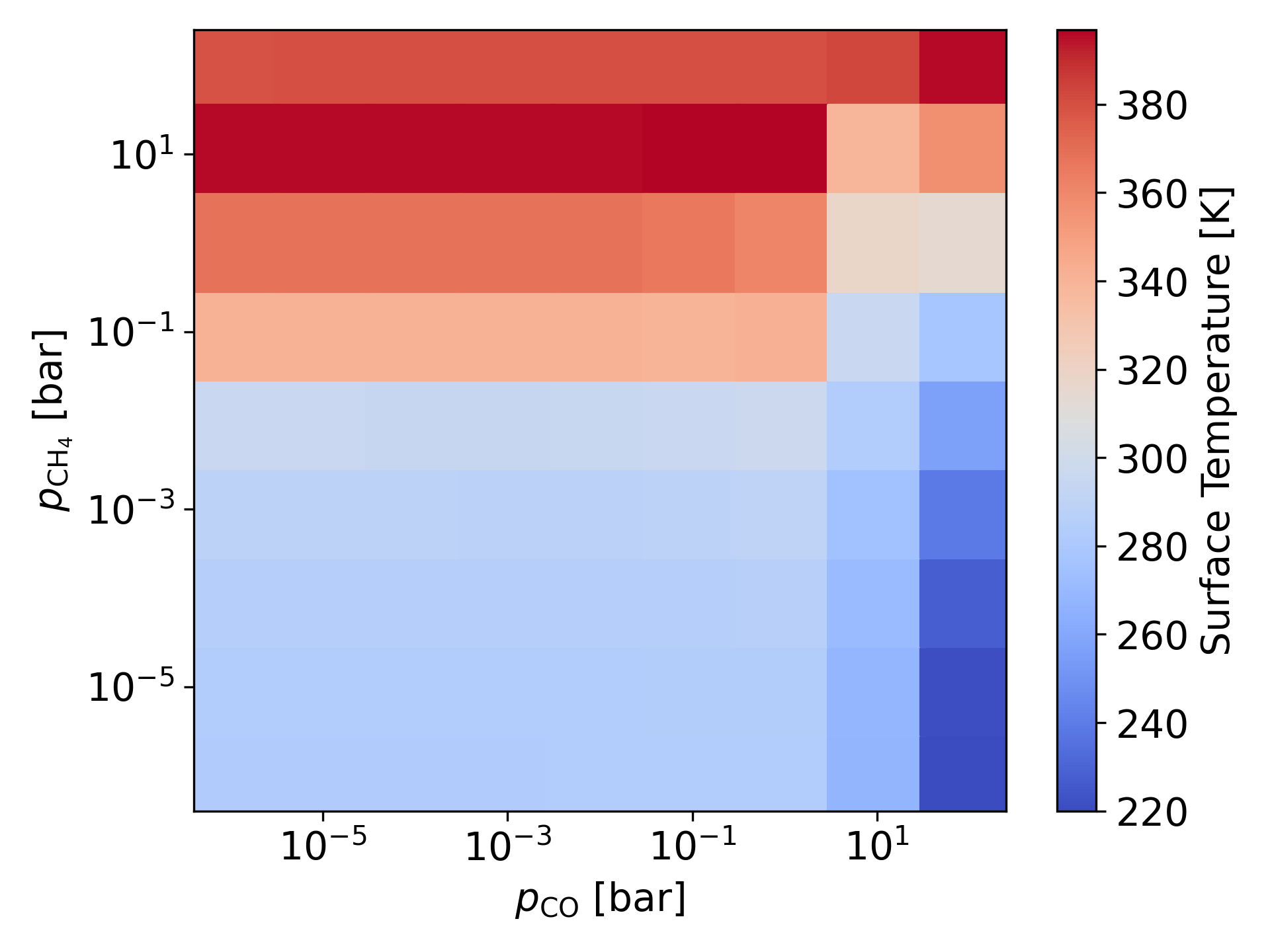}
        \put(-235,180){\scriptsize \textbf{(a)}}
        \label{fig:results:COCH4:panel:temp}
    \end{minipage}
    \hfill
    \begin{minipage}[b]{0.49\textwidth}
        \centering
        \includegraphics[width=\textwidth]{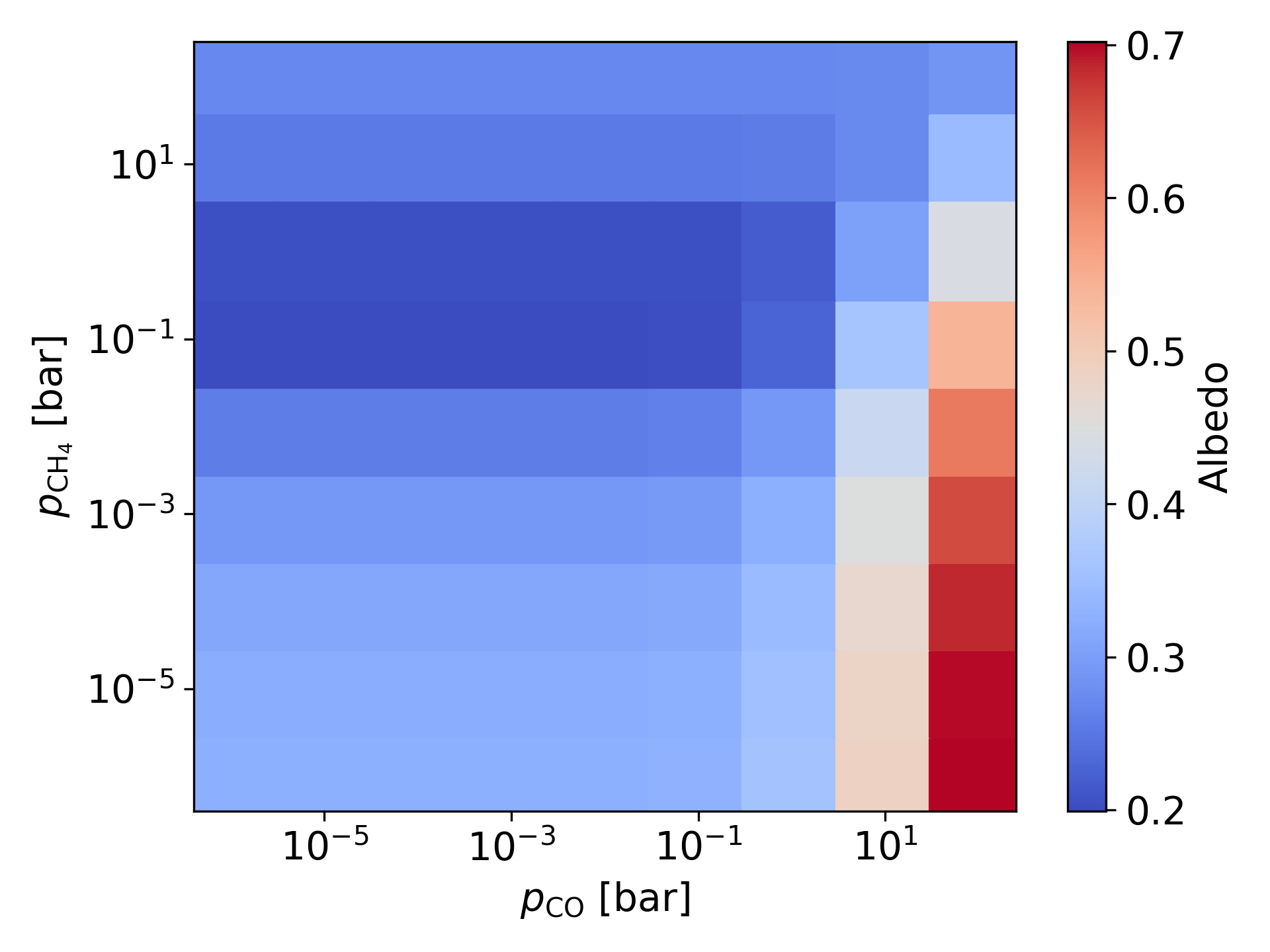}
        \put(-235,180){\scriptsize \textbf{(b)}}
        \label{fig:results:COCH4:panel:alb}
    \end{minipage}
    \begin{minipage}[b]{0.49\textwidth}
        \centering
        \includegraphics[width=\textwidth]{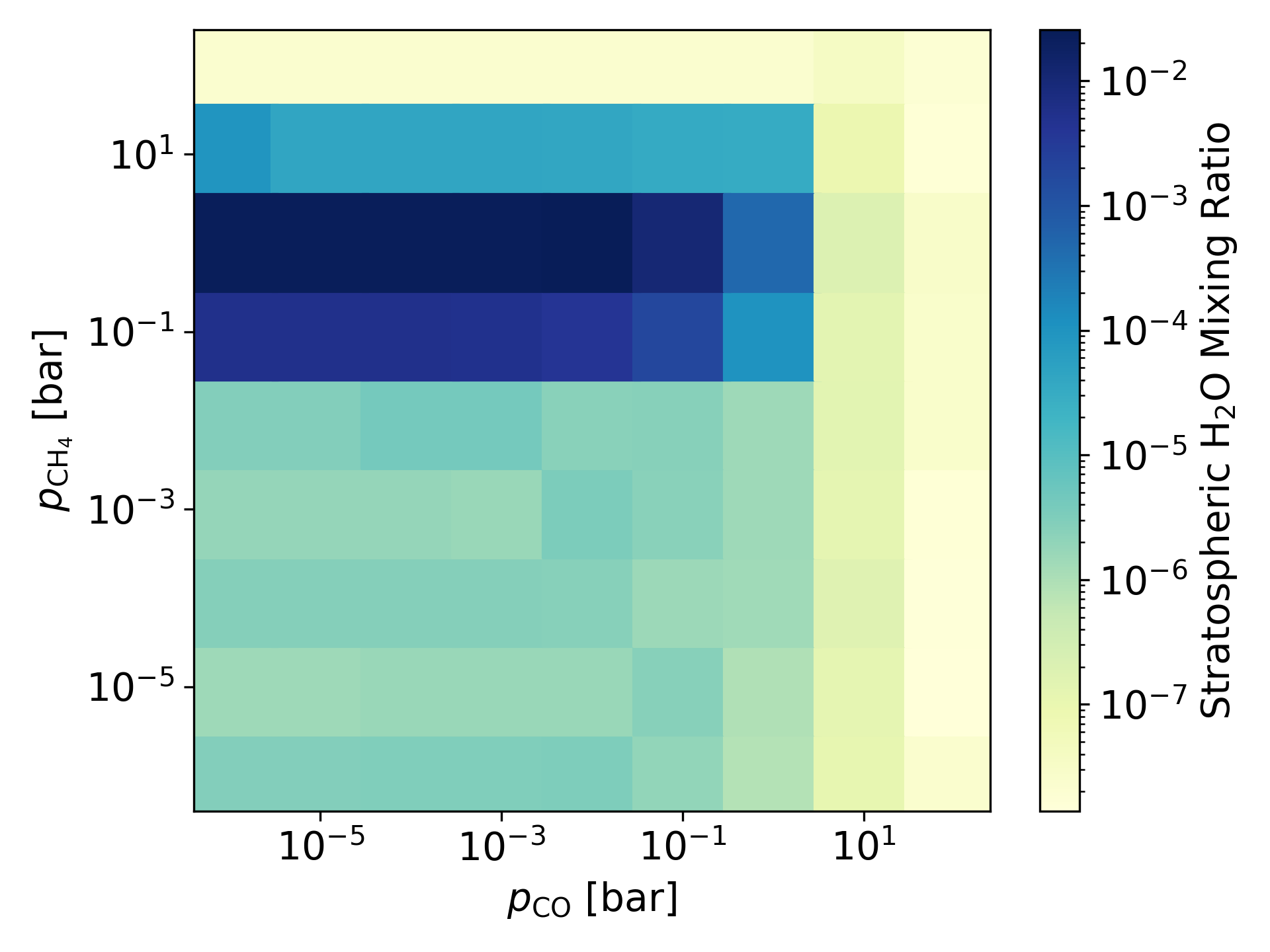}
        \put(-235,180){\scriptsize \textbf{(c)}}
        \label{fig:results:COCH4:panel:wv}
    \end{minipage}
    \hfill
    \begin{minipage}[b]{0.49\textwidth}
        \centering
        \includegraphics[width=\textwidth]{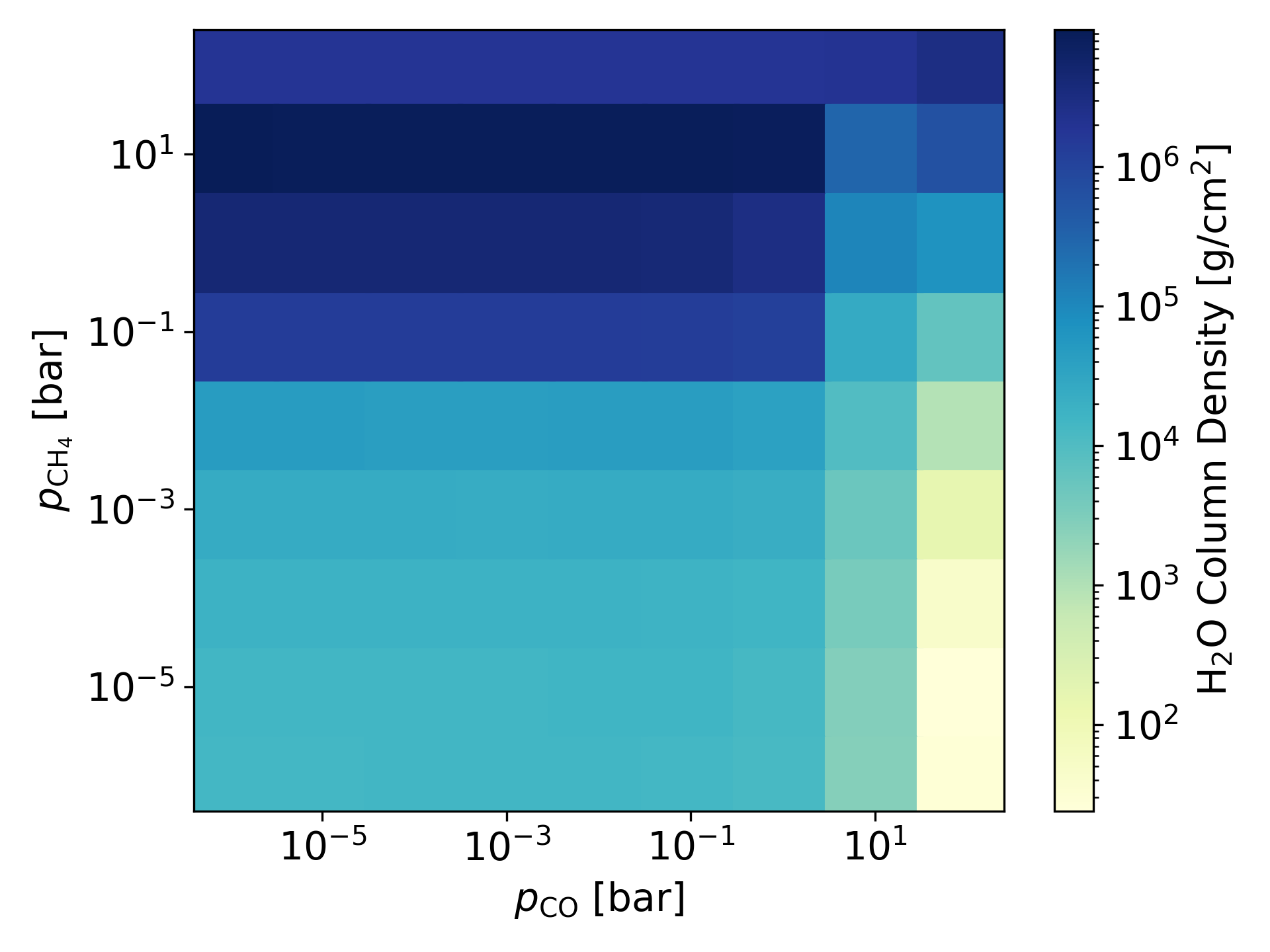}
        \put(-235,180){\scriptsize \textbf{(d)}}
        \label{fig:results:COCH4:panel:coldens}
    \end{minipage}

    \caption{\footnotesize Dependence of (a) surface temperature, (b) planetary albedo (the Bond albedo measured at the top of the atmosphere), (c) H\(_2\)O mixing ratio in the stratosphere, and (d) H\(_2\)O column density on the partial pressures of CO and CH\(_4\), for a planet orbiting the Sun with $p_\mathrm{CO_2} = 0.01\,\mathrm{bar}$.}
    \label{fig:results:COCH4}
\end{figure}

The amounts of individual carbon species control the climate of terrestrial planets when all other factors remain constant. Here, we show results with fixed $p_\mathrm{CO_2}$ to mimic a planet with active carbonate-silicate cycling, and vary $p_\mathrm{CO}$ and $p_\mathrm{CH_4}$ (Section \ref{subsec:methods:input}). We assumed the host star to be the Sun.

Increasing CO causes cooling for planets orbiting G-type stars, except for high $p_\mathrm{CH_4}$ and $p_\mathrm{CO_2}$ cases ($\gtrsim 1\ \mathrm{bar}$). Figure \ref{fig:results:COCH4} shows the dependencies of surface temperature, planetary albedo (the Bond albedo, defined as the ratio of scattered/reflected outgoing and incoming fluxes of stellar light measured at the top of the atmosphere), stratospheric water vapor mixing ratio, and water vapor column density on $p_\mathrm{CO}$ and $p_\mathrm{CH_4}$, with a fixed $p_\mathrm{CO_2} = 0.01\ \mathrm{bar}$. When $p_\mathrm{CH_4} \lesssim 1\ \mathrm{bar}$, we found that increasing $p_\mathrm{CO}$ causes significant cooling at $p_\mathrm{CO} \gtrsim 1\ \mathrm{bar}$, which exceeds background $p_\mathrm{N_2}$ (Figure \ref{fig:results:COCH4}a). This is because CO itself is not a greenhouse gas and is instead a scatterer \citep{catling2017atmospheric}. The Rayleigh scattering is responsible for the decrease in temperature by increasing planetary albedo (Figure \ref{fig:results:COCH4}b). 

However, when $p_\mathrm{CH_4} \gtrsim 1\ \mathrm{bar}$, the effect of CO gradually shifts to warming of the surface (Figure \ref{fig:results:COCH4}a). As previously reported for N\textsubscript{2}, the warming can be explained by pressure broadening of absorption lines of greenhouse gases \citep{Goldblatt+2009,wordsworth2013water}, combined with additional contributions from the greenhouse effect of water vapor \citep{Goldblatt+2009}. Higher pressure induces frequent collisions, and consequently, broadening of the absorption lines of the other greenhouse gases (CO\textsubscript{2}, CH\textsubscript{4}, and H\textsubscript{2}O), leading to efficient absorption of thermal radiation \citep{catling2017atmospheric}. The warming due to CO is also attributed in part due to the associated change in the water vapor content (Figure \ref{fig:results:COCH4}d).

At high CH\textsubscript{4} levels ($p_\mathrm{CH_4}$ \(\gtrsim\) 10 bar), the surface temperature begins to decrease with increasing $p_\mathrm{CH_4}$ (Figure \ref{fig:results:COCH4}a). This is attributed to the decline of another greenhouse gas, H\textsubscript{2}O ( Figures \ref{fig:results:COCH4}c and \ref{fig:results:COCH4}d). The surface temperature decrease following that of H\textsubscript{2}O content with increasing a greenhouse gas has been reported for CO\textsubscript{2} \citep{kasting1993earth,wordsworth2013water,ramirez2014can}. 
In this regime, increasing $p_\mathrm{CO_2}$ lowers the mixing ratio of water vapor at the surface. Consequently, the sensible heat of non-condensible gases exceeds the latent heat of water vapor, and the entire atmosphere up to the stratosphere becomes dry \citep{wordsworth2013water}. This explanation is applicable for CH\textsubscript{4} as well; at high CO\textsubscript{2} or CH\textsubscript{4} partial pressure, water vapor levels decrease in the stratosphere (Figure \ref{fig:results:COCH4}c), as well as total water vapor levels decreasing in the atmosphere (Figure \ref{fig:results:COCH4}d). The surface temperature decreases from a diminished greenhouse effect combined with a higher planetary albedo due to scattering (Figure \ref{fig:results:COCH4}b). We note that, in Figure \ref{fig:results:COCH4}, there is a region in which the stratospheric water vapor mixing ratio and the column density of water vapor do not correlate with each other. When this occurs, it is because the total atmospheric pressure also changes and the increase of (in this case)  $p_\mathrm{CO}$ is larger than that of water vapor. Thus, while the total amount of water vapor in the column increases when CO causes warming, the relative increase of water vapor compared to CO is actually lower (Figs \ref{fig:results:COCH4}c and \ref{fig:results:COCH4}d).

Furthermore, the influence of CH\textsubscript{4} has slight changes depending on the partial pressure of CO. When $p_\mathrm{CO}\gtrsim 10\ \mathrm{bar}$, the addition of methane will cause warming from as low as $p_\mathrm{CH_4} \sim 10^{-6}\ \mathrm{bar}$, and no longer causes cooling (Figure \ref{fig:results:COCH4}a). At high $p_\mathrm{CH_4}$, we find that the addition of CH\textsubscript{4} causes warming instead of cooling as it did when $p_\mathrm{CO}$ was low. In this region, the pressure broadening from CO offsets the cooling effect of CH\textsubscript{4}. However, the overall behavior of CH\textsubscript{4} does not change.

As was discussed briefly earlier in Section \ref{sec:intro}, organic haze might form and cools the surface at $f_\mathrm{CH_4}/f_\mathrm{CO_2}\gtrsim 0.1$ \citep{pavlov2000greenhouse,haqq2008revised}. Thus, the surface temperature shown Figure \ref{fig:results:COCH4}a needs to be regarded as an upper limit, and the cooling with increasing $p_\mathrm{CH_4}$ can start around $p_\mathrm{CH_4}\sim 10^{-3}\ \mathrm{bar}$.
We revisit this issue in Section \ref{sec:Discussion}.

\subsection{\texorpdfstring{Dependence on $CO_\mathrm{2}$ Partial Pressure}{Dependence on CO2 Partial Pressure}}\label{subsec:results:CO2}

\setlength{\abovecaptionskip}{0pt}
\setlength{\belowcaptionskip}{0pt}

\begin{figure}[ht]
    \centering
    \begin{minipage}[b]{0.49\textwidth}
        \centering
        \includegraphics[width=\textwidth]{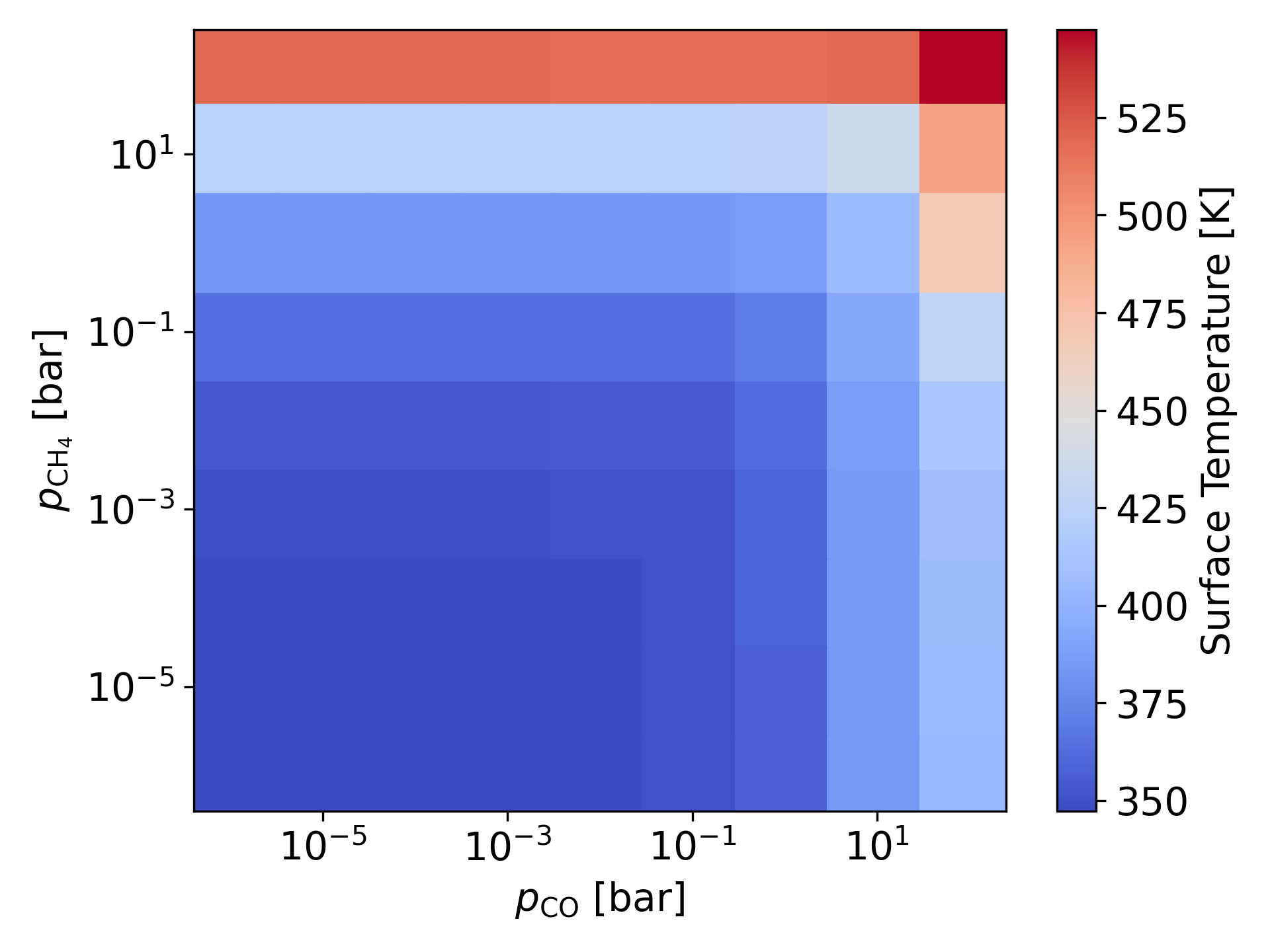}
        \put(-235,180){\scriptsize \textbf{(a)}}
        \label{fig:results:highCO2:panel:temp}
    \end{minipage}
    \hfill
    \begin{minipage}[b]{0.49\textwidth}
        \centering
        \includegraphics[width=\textwidth]{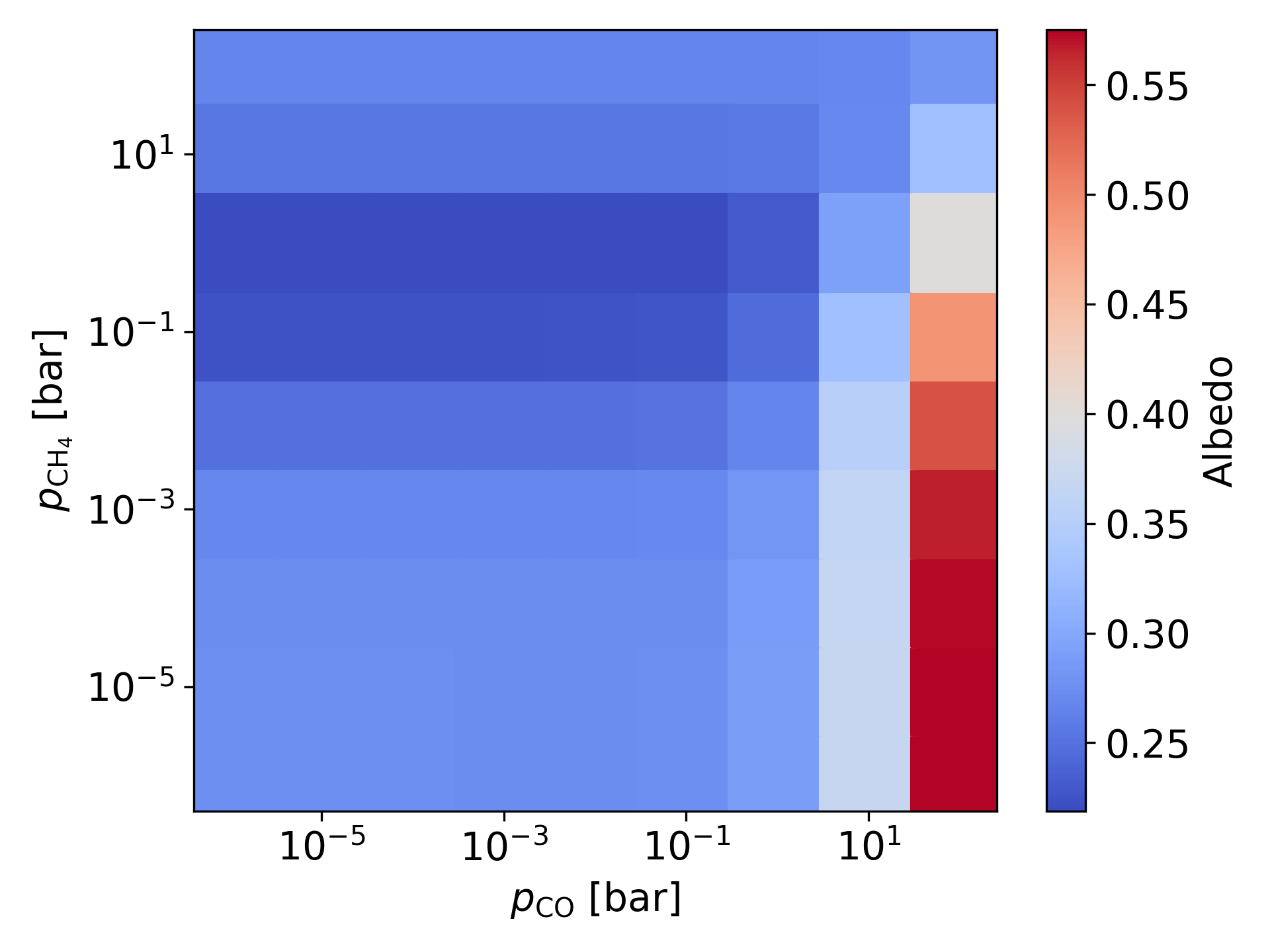}
        \put(-235,180){\scriptsize \textbf{(b)}}
        \label{fig:results:highCO2:panel:alb}
    \end{minipage}
    \begin{minipage}[b]{0.49\textwidth}
        \centering
        \includegraphics[width=\textwidth]{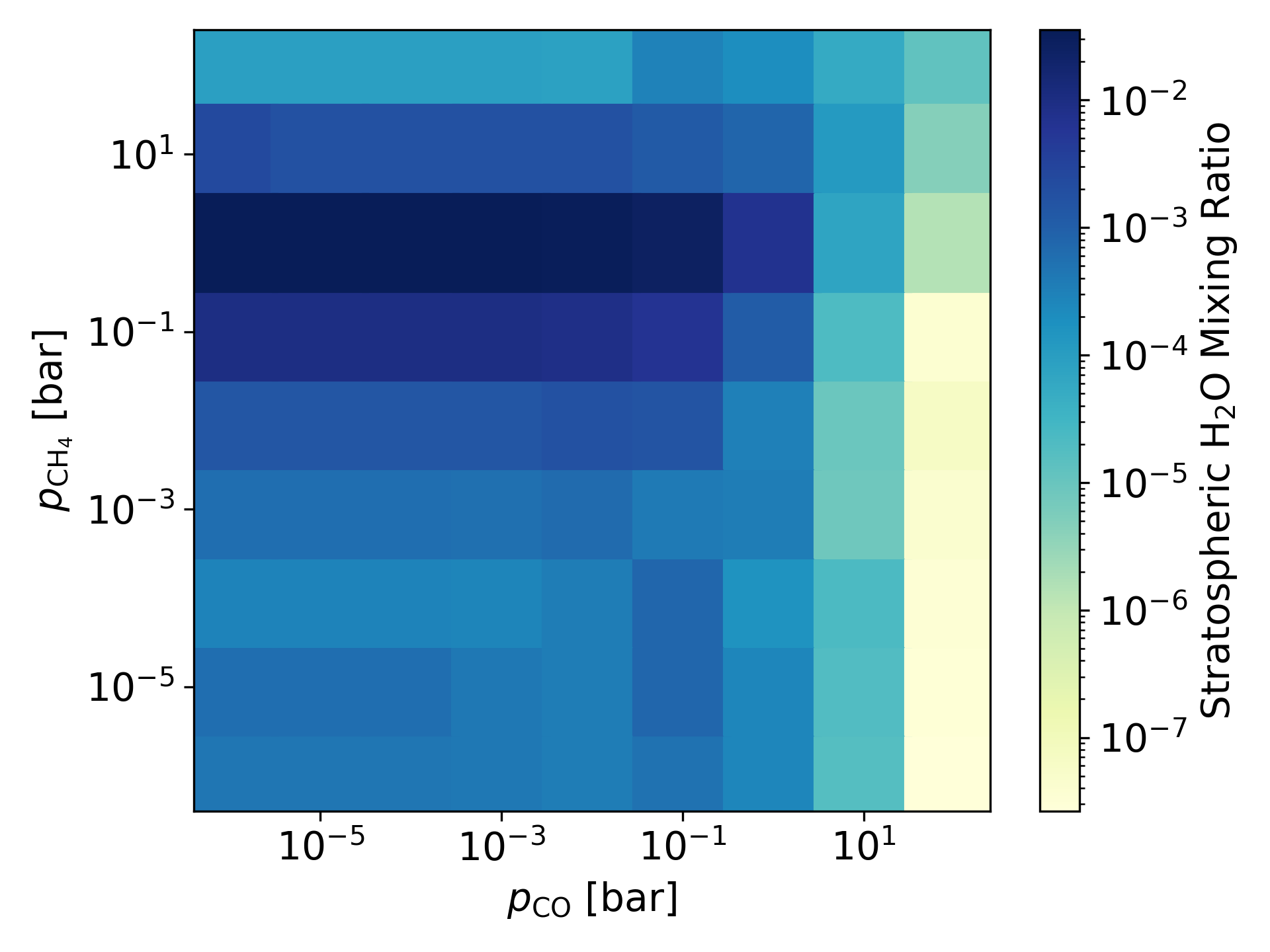}
        \put(-235,180){\scriptsize \textbf{(c)}}
        \label{fig:results:highCO2:panel:wv}
    \end{minipage}
    \hfill
    \begin{minipage}[b]{0.49\textwidth}
        \centering
        \includegraphics[width=\textwidth]{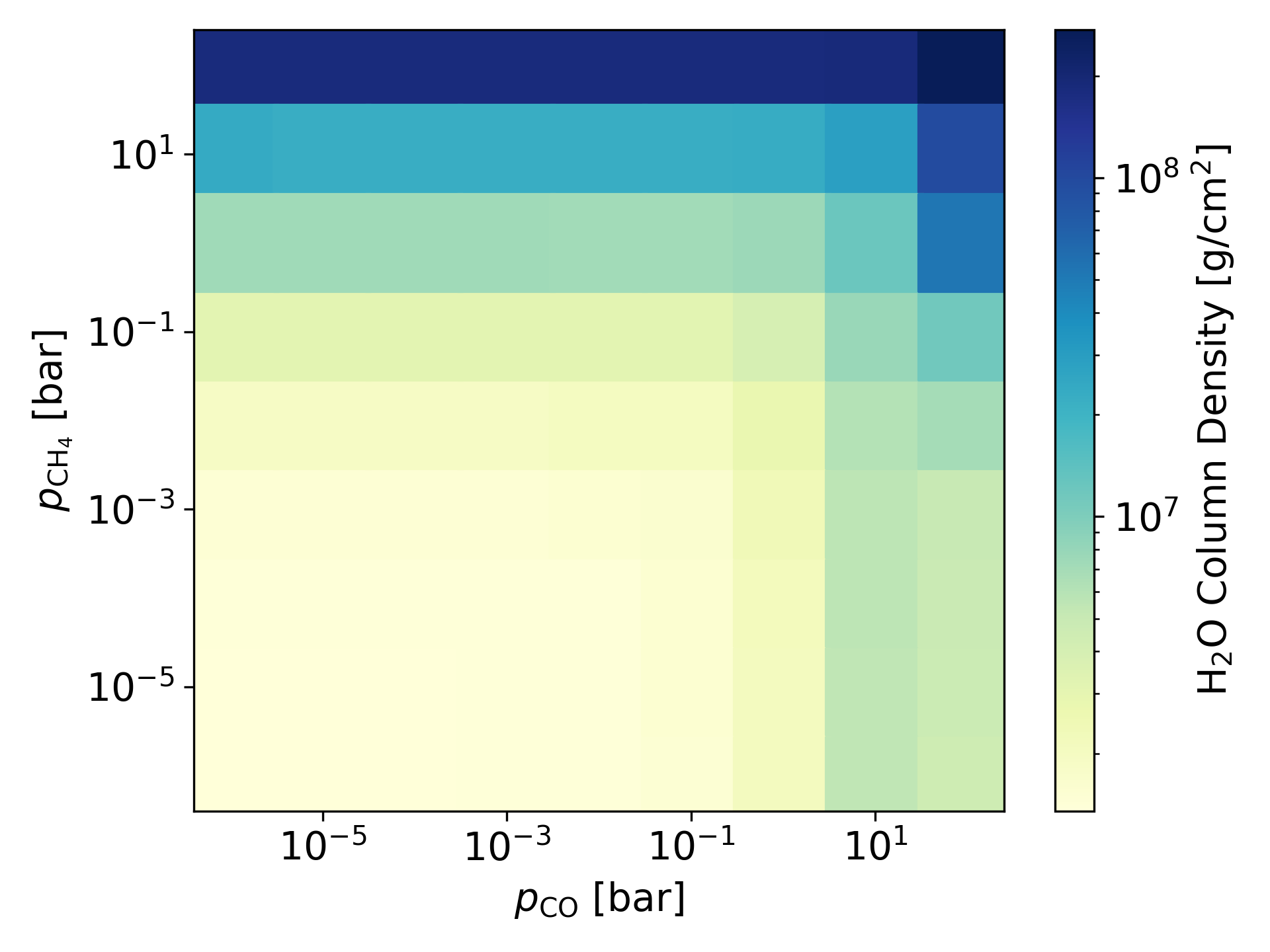}
        \put(-235,180){\scriptsize \textbf{(d)}}
        \label{fig:results:highCO2:panel:coldens}
    \end{minipage}

    \caption{\footnotesize Same as Figure \ref{fig:results:COCH4} but showing the results with $p_\mathrm{CO_2} = 1\ \mathrm{bar}$.}
    \label{fig:results:highCO2}
\end{figure}

With higher levels of CO\textsubscript{2}, the parameter space where increasing CO and CH\textsubscript{4} causes warming expands. We show results with $p_\mathrm{CO_2} = 1\ \mathrm{bar}$ in Figure \ref{fig:results:highCO2}. The overall temperature of the atmosphere across the parameter space increases due to an increased greenhouse effect from CO\textsubscript{2} and the corresponding increase in water vapor levels (Figs \ref{fig:results:highCO2}a and \ref{fig:results:highCO2}d), compared to the $p_\mathrm{CO_2} = 0.01\ \mathrm{bar}$ case (Figure \ref{fig:results:COCH4}a). In contrast to the low $p_\mathrm{CO_2}$ case, increasing CO always leads to warming in the $p_\mathrm{CO}$ and $p_\mathrm{CH_4}$ parameter space studied here, suggesting that the transition from cooling to warming depends both on $p_\mathrm{CO_2}$ and $p_\mathrm{CH_4}$. 

The transition to warming is believed to be driven by the dominance of pressure broadening and increased H\textsubscript{2}O content (Figure \ref{fig:results:highCO2}d) over the increase in albedo due to scattering (Figure \ref{fig:results:highCO2}b). We also found that increasing $p_\mathrm{CO_2}$ from 0.01 bar to 1 bar has a greater impact on the transition between CO cooling and warming than changing $p_\mathrm{CH_4}$ by the same magnitude (Figures \ref{fig:results:COCH4}a and \ref{fig:results:highCO2}a). Because CO\textsubscript{2} is a more potent greenhouse gas than CH\textsubscript{4}, the transition is more dependent on changes in CO\textsubscript{2} than it is on CH\textsubscript{4}. 

Cooling with increasing CH\textsubscript{4} at high $p_\mathrm{CH_4}$ is no longer present in the $p_\mathrm{CO_2} = 1\ \mathrm{bar}$ case (Figure \ref{fig:results:highCO2}a). Carbon dioxide dominates the greenhouse effect in this case, and prevents inhibition of H\textsubscript{2}O from CH\textsubscript{4}. Above 10 bar $p_\mathrm{CH_4}$, albedo decreases with increasing $p_\mathrm{CH_4}$ (Figure \ref{fig:results:highCO2}b). H\textsubscript{2}O decreases in the stratosphere but not the entire atmosphere (Figures \ref{fig:results:highCO2}c and \ref{fig:results:highCO2}d). Therefore, as was the case with the $p_\mathrm{CH_4}$ \(<\) 10 bar in the low $p_\mathrm{CO_2}$ case (Figure \ref{fig:results:COCH4}a), increasing CH\textsubscript{4} results in an increase in the surface temperature even for higher $p_\mathrm{CH_4}$.

\subsection{Dependence on Spectral Type of the Host Star}\label{subsec:results:Spec}

\setlength{\abovecaptionskip}{0pt}
\setlength{\belowcaptionskip}{0pt}

\begin{figure}[ht]
    \centering
    \begin{minipage}[b]{0.49\textwidth}
        \centering
        \includegraphics[width=\textwidth]{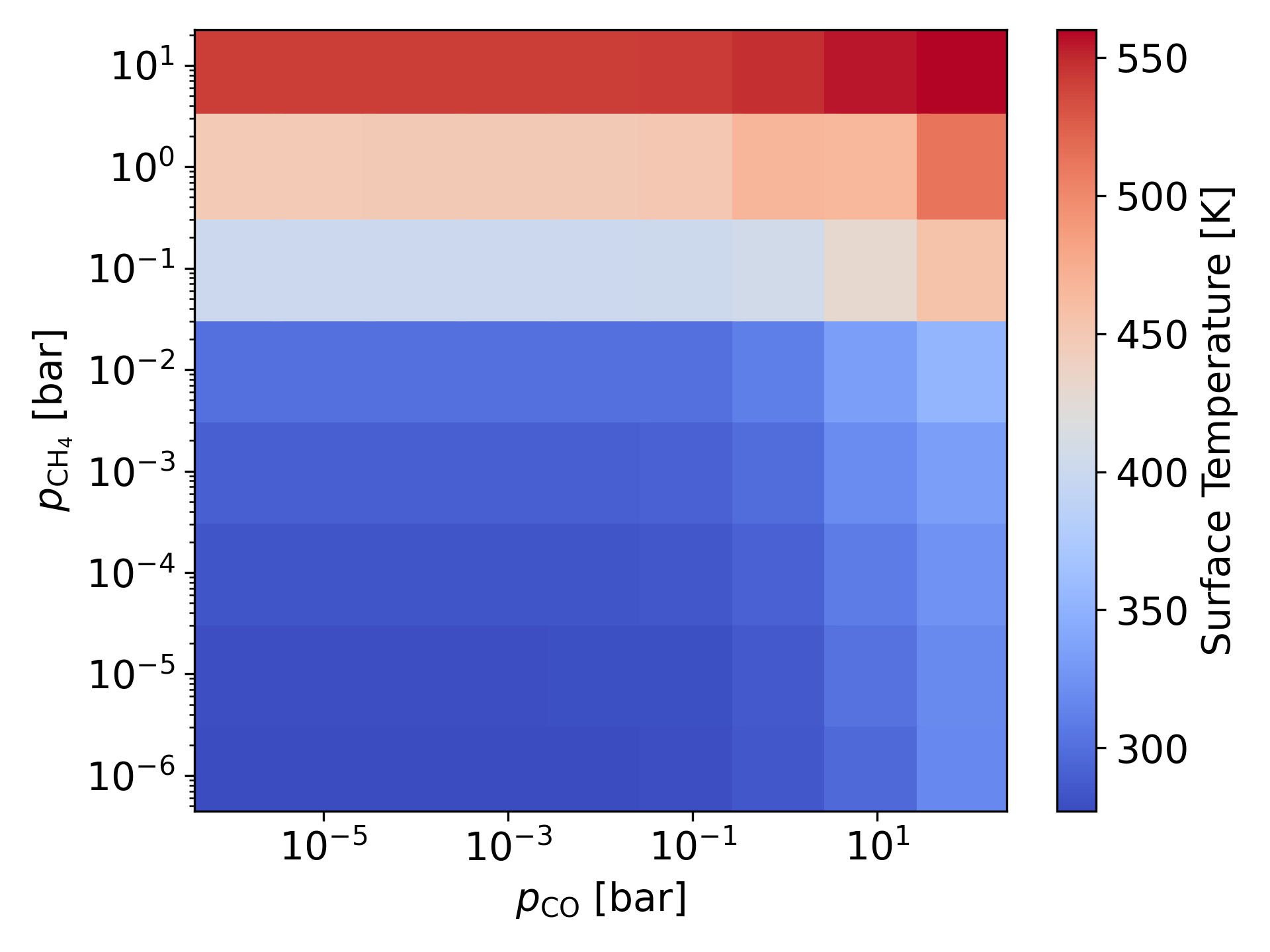}
        \put(-235,180){\scriptsize \textbf{(a)}}
        \label{fig:results:Mstar:panel:temp}
    \end{minipage}
    \hfill
    \begin{minipage}[b]{0.49\textwidth}
        \centering
        \includegraphics[width=\textwidth]{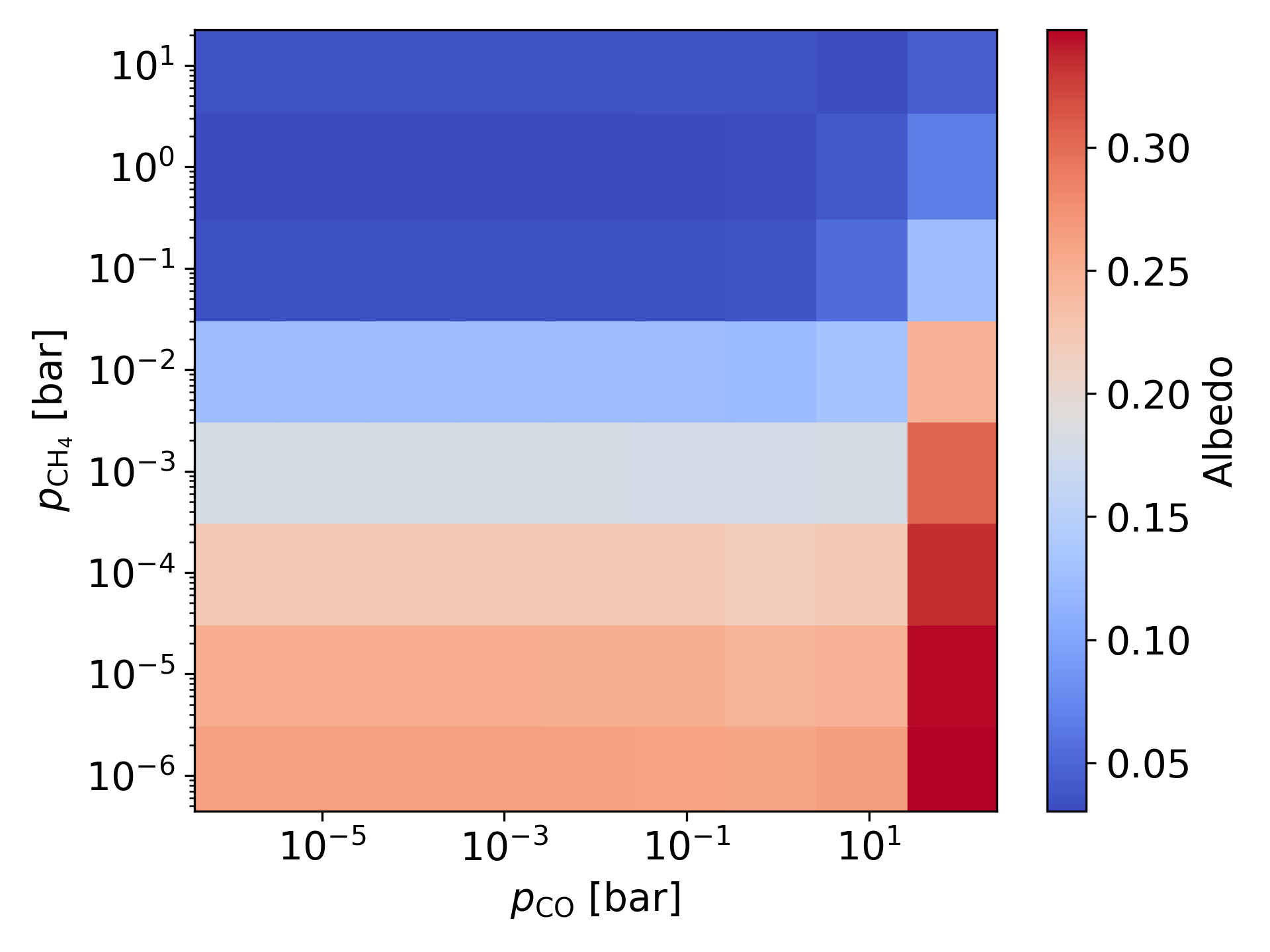}
        \put(-235,180){\scriptsize \textbf{(b)}}
        \label{fig:results:Mstar:panel:alb}
    \end{minipage}
    \begin{minipage}[b]{0.49\textwidth}
        \centering
        \includegraphics[width=\textwidth]{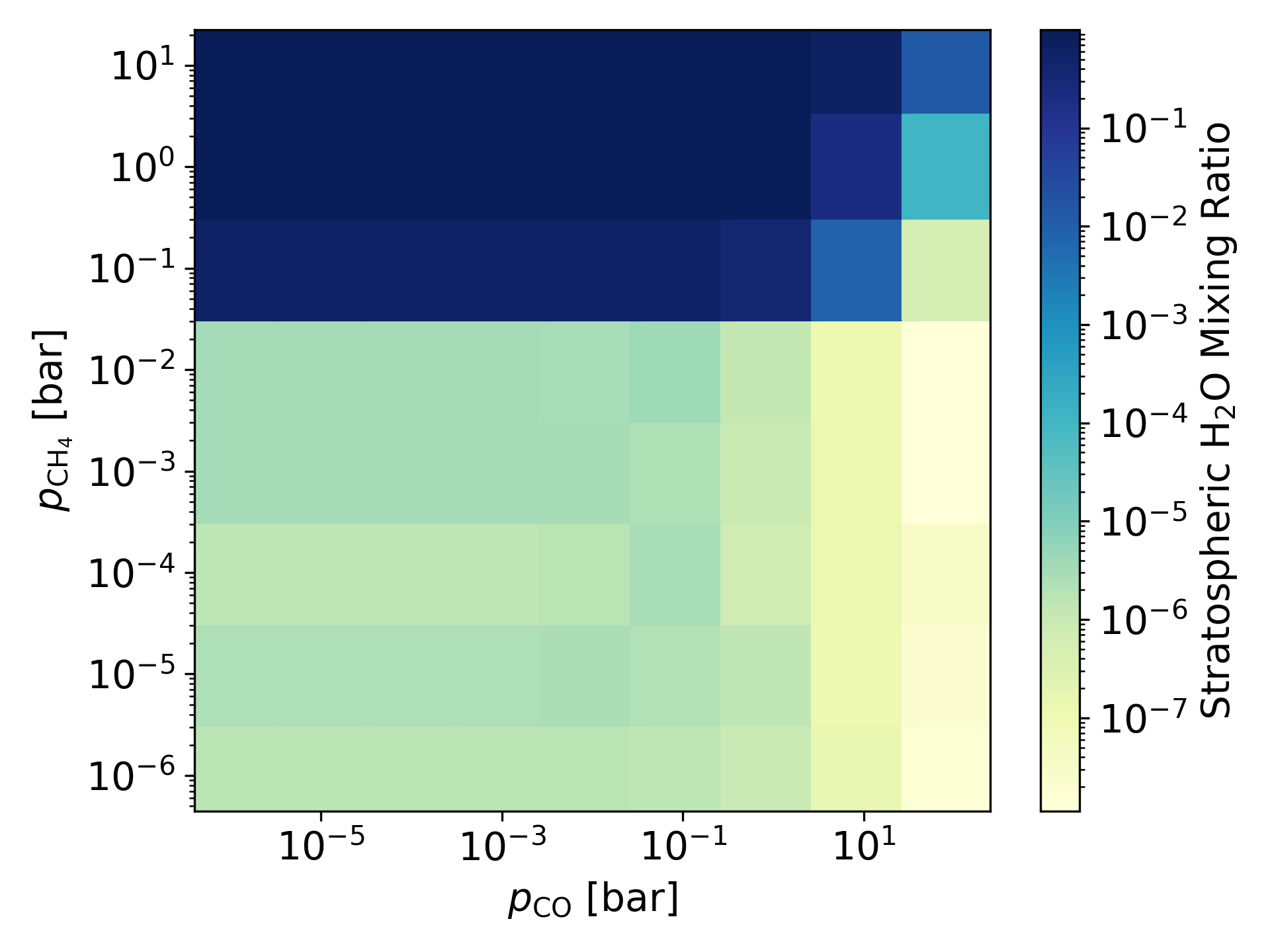}
        \put(-235,180){\scriptsize \textbf{(c)}}
        \label{fig:results:Mstar:panel:wv}
    \end{minipage}
    \hfill
    \begin{minipage}[b]{0.49\textwidth}
        \centering
        \includegraphics[width=\textwidth]{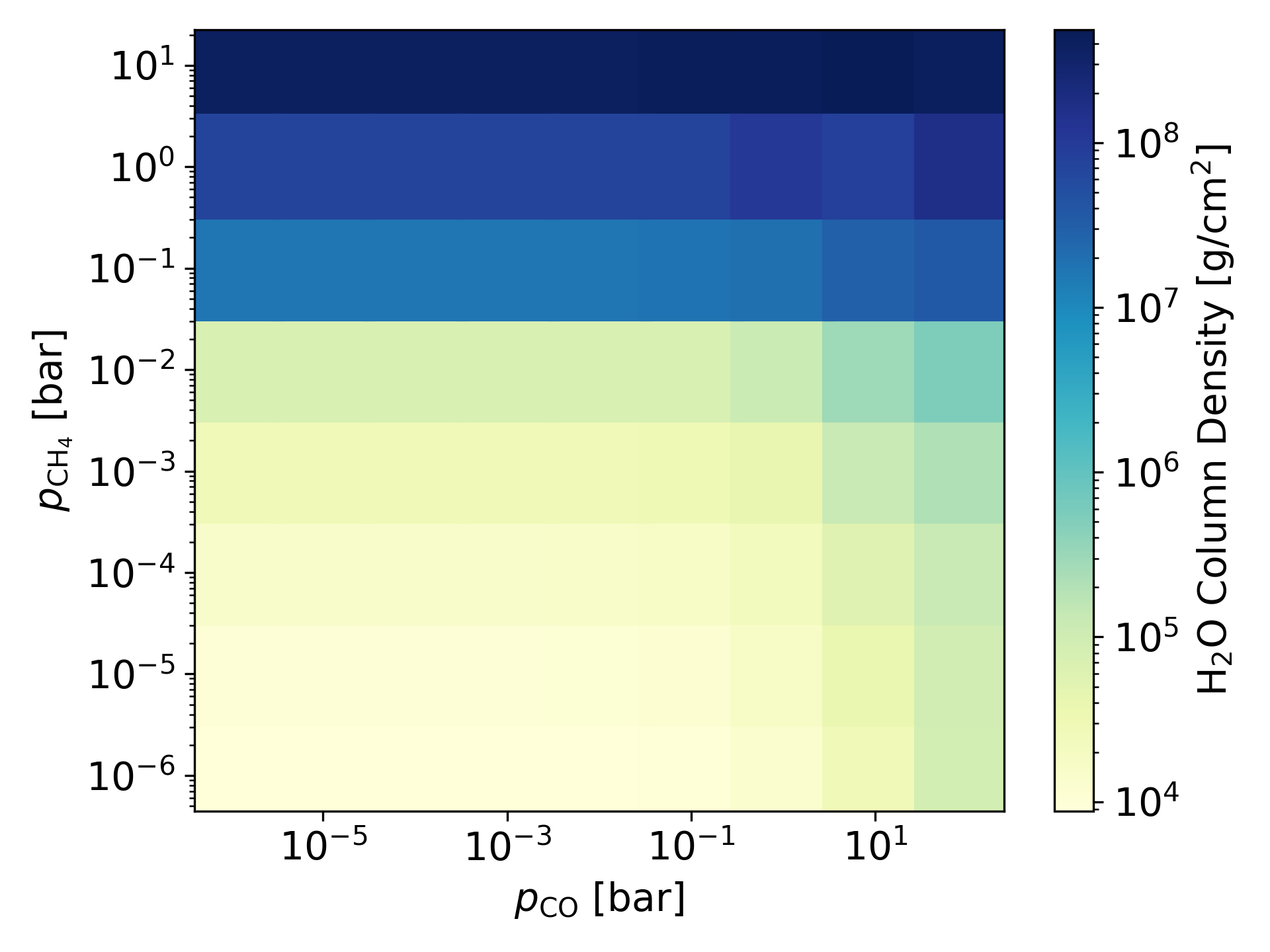}
        \put(-235,180){\scriptsize \textbf{(d)}}
        \label{fig:results:Mstar:panel:coldens}
    \end{minipage}

    \caption{\footnotesize Same as Figure \ref{fig:results:COCH4} but showing the results with GJ 876 being the host star.}
    \label{fig:results:Mstar}
\end{figure}

For planets orbiting M-type stars, warming with increasing CO is more dominant. Figure \ref{fig:results:Mstar} shows the results with the host star being the M-dwarf GJ 876 and with the fiducial $p_\mathrm{CO_2} = 0.01\ \mathrm{bar}$. Even at this lower $p_\mathrm{CO_2}$ level, CO acts as an indirect greenhouse gas, especially when $p_\mathrm{CO} \gtrsim 1\ \mathrm{bar}$ (namely, $p_\mathrm{CO}>p_\mathrm{N_2}$; Figure \ref{fig:results:Mstar}a). The radiation from M-type stars is weaker than G-types in the visible range \citep[Figure \ref{fig:methods:spectra},][]{harman2015abiotic,arney2017pale}. Since the Rayleigh scattering cross section is inversely proportional to the fourth power of wavelength \citep{catling2017atmospheric}, Rayleigh scattering by CO becomes weaker compared to the case of a G-type host star, as indicated by the limited increase in albedo with increasing $p_\mathrm{CO}$ (Figure \ref{fig:results:Mstar}b vs. Figure \ref{fig:results:COCH4}b). However, the indirect warming by pressure broadening and the corresponding high water vapor levels remain unchanged. Thus, warming dominates over cooling.
 
In terms of CH\textsubscript{4}, increasing $p_\mathrm{CH_4}$ always causes warming in the $p_\mathrm{CO_2}$ and $p_\mathrm{CH_4}$ parameter space studied (Figure \ref{fig:results:Mstar}a). This result is in contrast to the case of a G-type host star (Figure \ref{fig:results:COCH4}a), where cooling with increasing $p_\mathrm{CH_4}$ was seen for $p_\mathrm{CH_4}\gtrsim 10\ \mathrm{bar}$. Such contrasting results between G- and M-type host stars have been reported again for CO\textsubscript{2}, as shown in \cite{wordsworth2013water} (their Figure 7 vs. Figure 11). Reduced Rayleigh scattering and absorption of stellar light by CH\textsubscript{4} inhibit the increase in albedo even for the high $p_\mathrm{CH_4}$ (Figure \ref{fig:results:Mstar}b). The resultant higher temperature results in higher stratospheric mixing ratios and H\textsubscript{2}O column densities (Figures \ref{fig:results:Mstar}c and \ref{fig:results:Mstar}d).

\subsection{Case of Fixed Total Carbon Content}\label{subsec:results:pCtotal}

\setlength{\abovecaptionskip}{0pt}
\setlength{\belowcaptionskip}{0pt}

\begin{figure}[ht]
    \centering
    \begin{minipage}[b]{0.49\textwidth}
        \centering
        \includegraphics[width=\textwidth]{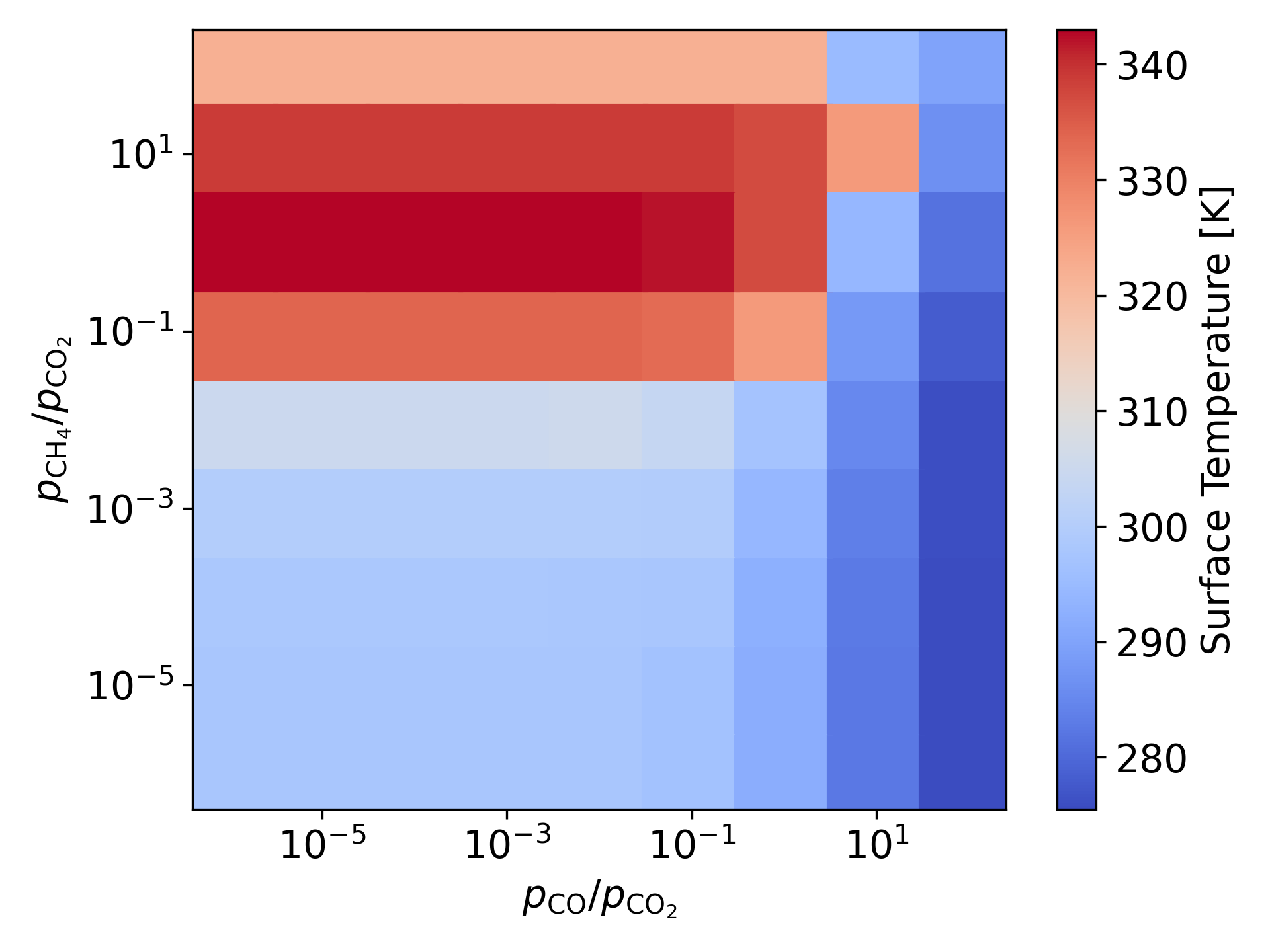}
        \put(-235,180){\scriptsize \textbf{(a)}}
        \label{fig:results:Fixed_figure:panel:temp}
    \end{minipage}
    \hfill
    \begin{minipage}[b]{0.49\textwidth}
        \centering
        \includegraphics[width=\textwidth]{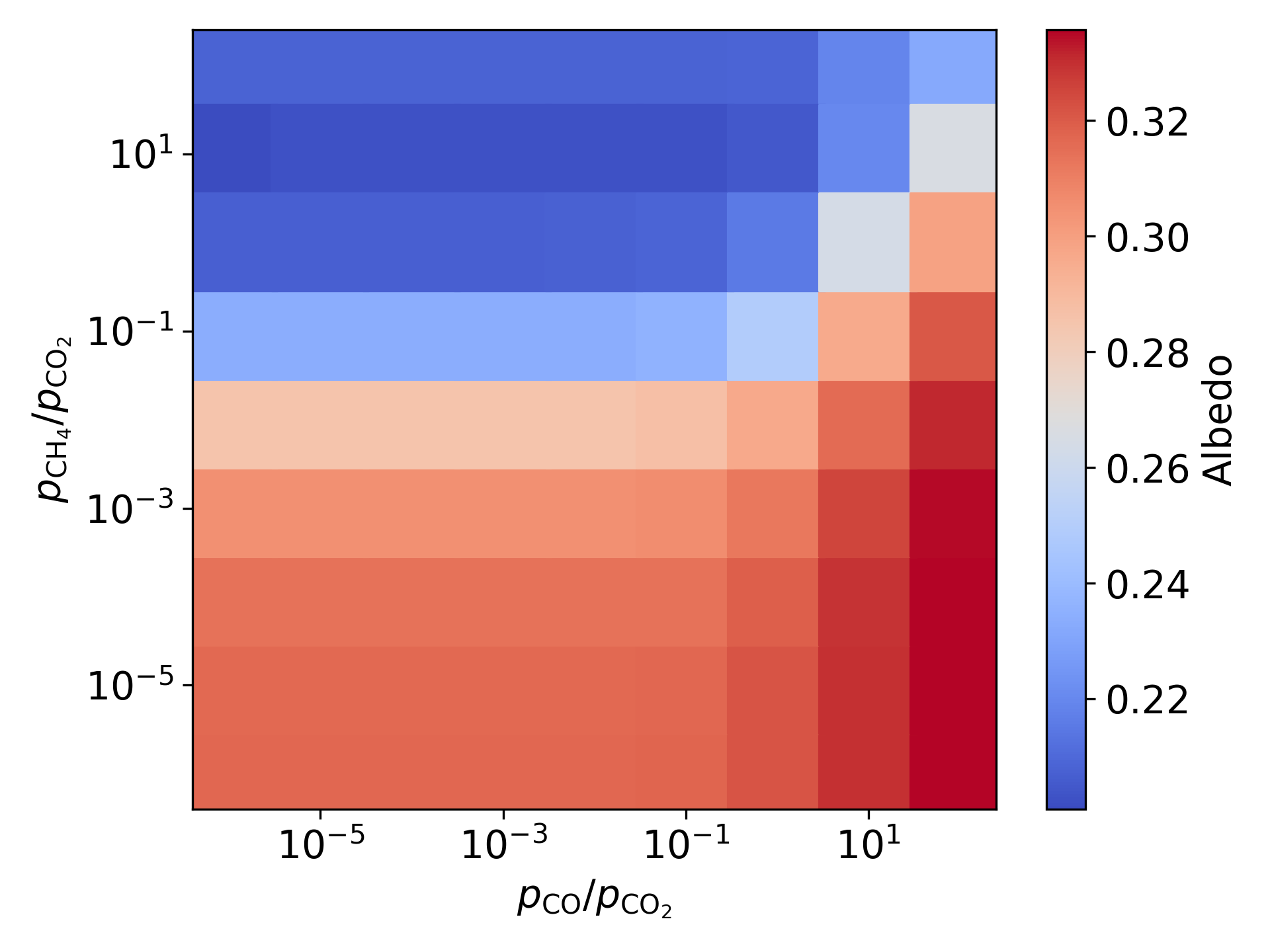}
        \put(-235,180){\scriptsize \textbf{(b)}}
        \label{fig:results:Fixed_figure:panel:alb}
    \end{minipage}
    \begin{minipage}[b]{0.49\textwidth}
        \centering
        \includegraphics[width=\textwidth]{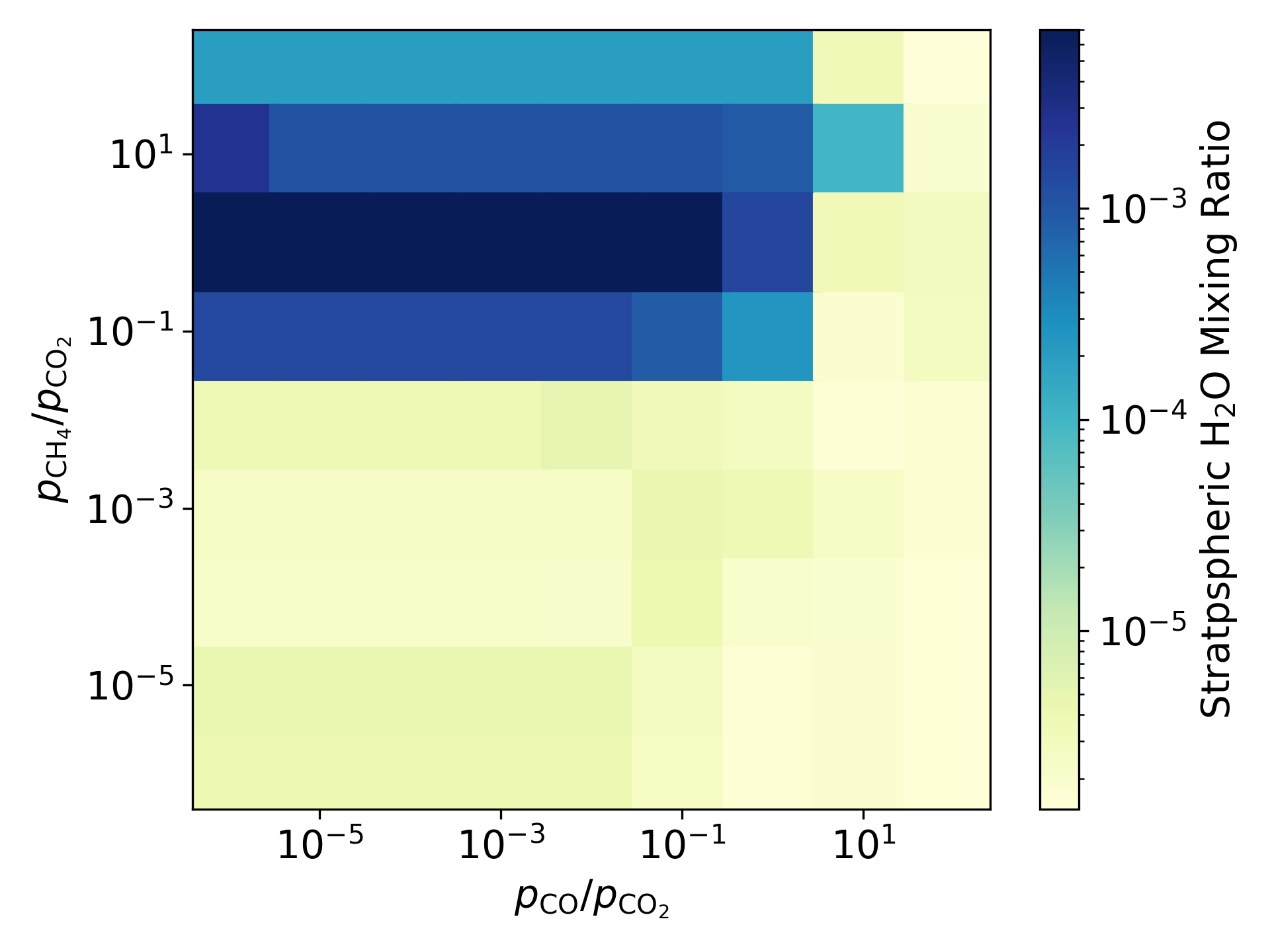}
        \put(-235,180){\scriptsize \textbf{(c)}}
        \label{fig:results:Fixed_figure:panel:wv}
    \end{minipage}
    \hfill
    \begin{minipage}[b]{0.49\textwidth}
        \centering
        \includegraphics[width=\textwidth]{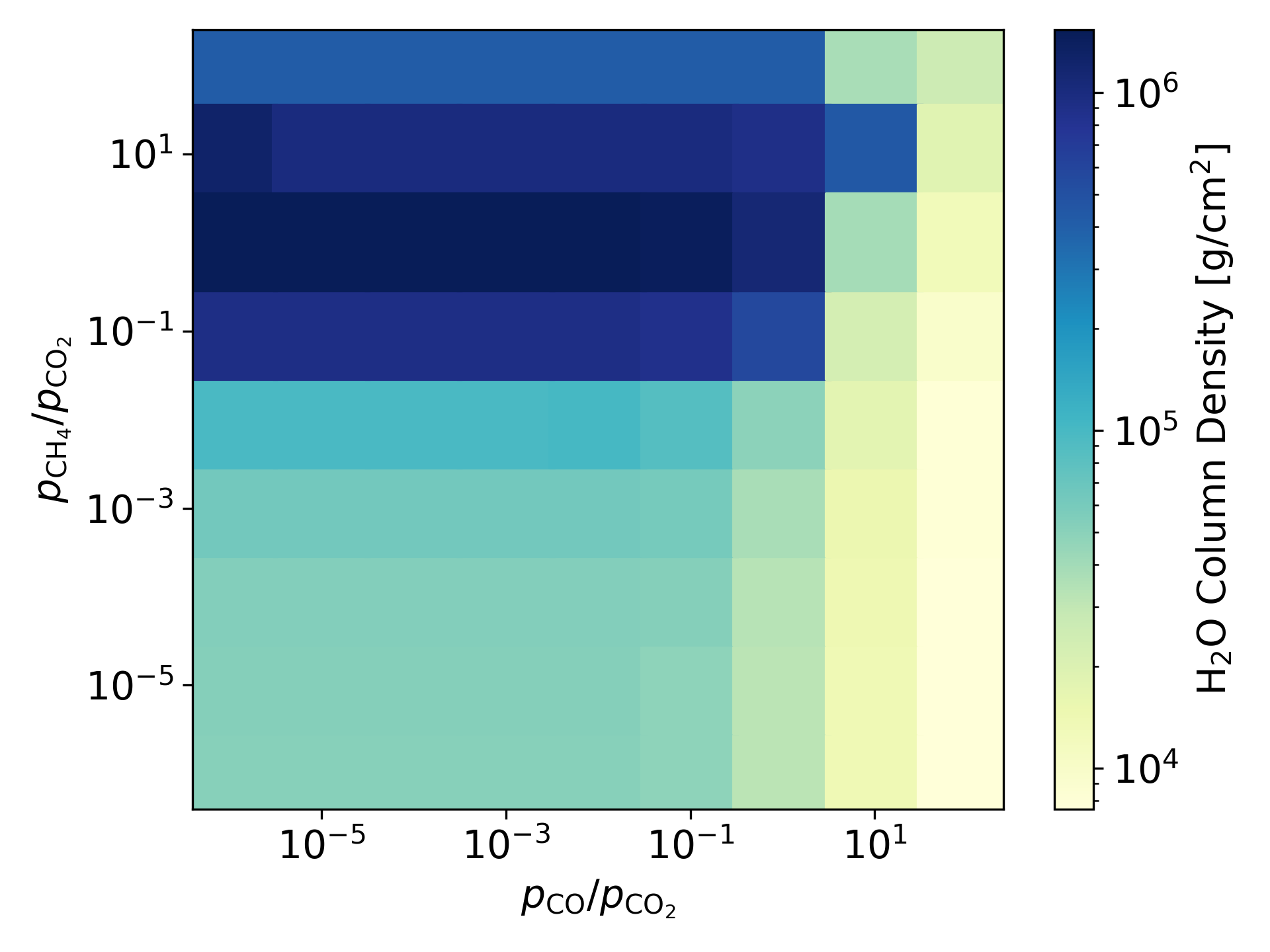}
        \put(-235,180){\scriptsize \textbf{(d)}}
        \label{fig:results:Fixed_figure:panel:coldens}
    \end{minipage}

    \caption{\footnotesize Same as Figure \ref{fig:results:COCH4} but showing the dependence on $p_\mathrm{CO}$/$p_\mathrm{CO_2}$ and $p_\mathrm{CH_4}$/$p_\mathrm{CO_2}$ with a fixed $p_\mathrm{Ctotal} = 0.1\ \mathrm{bar}$.}
    \label{fig:results:Fixed_figure}
\end{figure}

For situations in which the carbonate-silicate cycle is inactive, which is likely for planets without global tectonism, we adopt a model whereby the total carbon content of the atmosphere system does not change. This allows us to mimic systems in which there is no transfer into and out of the system for carbon and nitrogen, though we do not fix the total amount of hydrogen or oxygen. We found that increasing $p_\mathrm{CO}/p_\mathrm{CO_2}$ cools the surface (Figure \ref{fig:results:Fixed_figure}a), as expected from decreasing a greenhouse gas, CO\textsubscript{2}. The cooling is also attributed to CO Rayleigh scattering, as indicated by the increasing albedo (Figure \ref{fig:results:Fixed_figure}b). Pressure broadening cannot compensate for the cooling in this system, since the total pressure of the atmosphere is kept constant.

When we change $p_\mathrm{CH_4}$/$p_\mathrm{CO_2}$, the surface temperature peaks at $p_\mathrm{CH_4}$/$p_\mathrm{CO_2} \sim 1$ (Figure \ref{fig:results:Fixed_figure}a). 
These greenhouse gases have strong absorption bands in different wavelengths: for instance, $\simeq 13\ \mathrm{\mu m}$ and $\simeq 8\ \mathrm{\mu m}$ for CO\textsubscript{2} and CH\textsubscript{4}, respectively \citep{catling2017atmospheric}. Thus, mixing with roughly 1:1 ratio maximizes the net greenhouse effect, although the precise value will be dependent on the relative strength of their absorption and the surface temperature which determine the black body spectrum.   

Furthermore, we found that H\textsubscript{2}O levels, both stratospheric mixing ratios and column densities, peak at $p_\mathrm{CH_4}$/$p_\mathrm{CO_2}\sim 1$ (Figures \ref{fig:results:Fixed_figure}c and \ref{fig:results:Fixed_figure}d). Like the previous cases, the water vapor levels as a result of changes in temperature amplify the already existing warming or cooling effects.

\section{Discussion} \label{sec:Discussion}

\subsection{Climate-Photochemistry Feedback}\label{subsec:DiscussionFeedback}

\begin{figure}[ht]
    \centering
    % First image (a)
    \begin{minipage}[b]{0.49\textwidth}
        \centering
        \includegraphics[width=\textwidth]{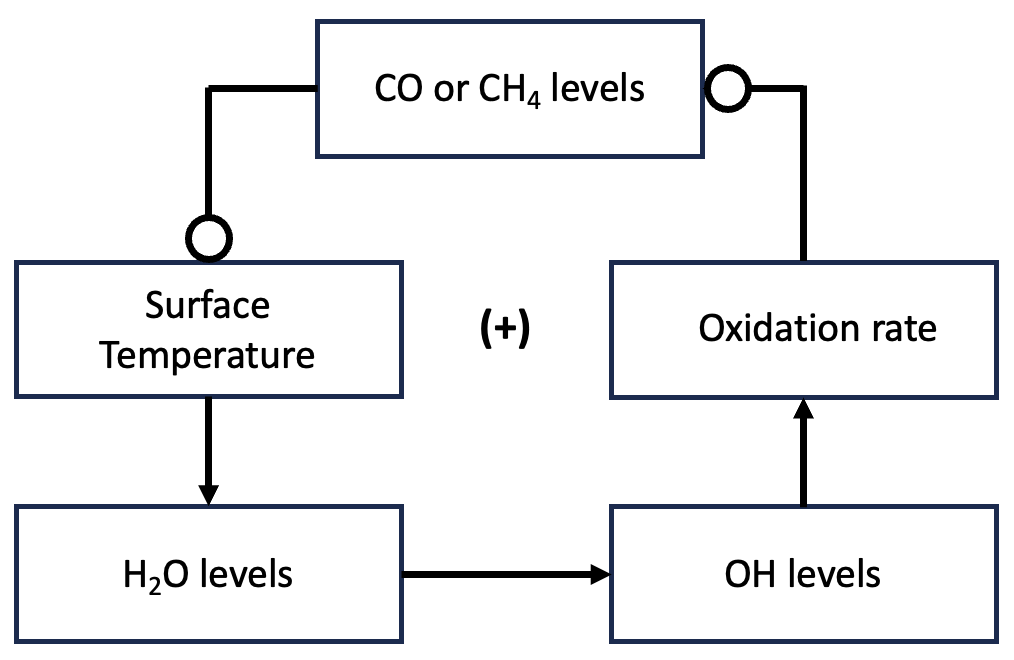}
        \put(-235,180){\scriptsize \textbf{(a)}}
        \label{fig:positive_feedback}
    \end{minipage}
    \hfill
    % Second image (b)
    \begin{minipage}[b]{0.49\textwidth}
        \centering
        \includegraphics[width=\textwidth]{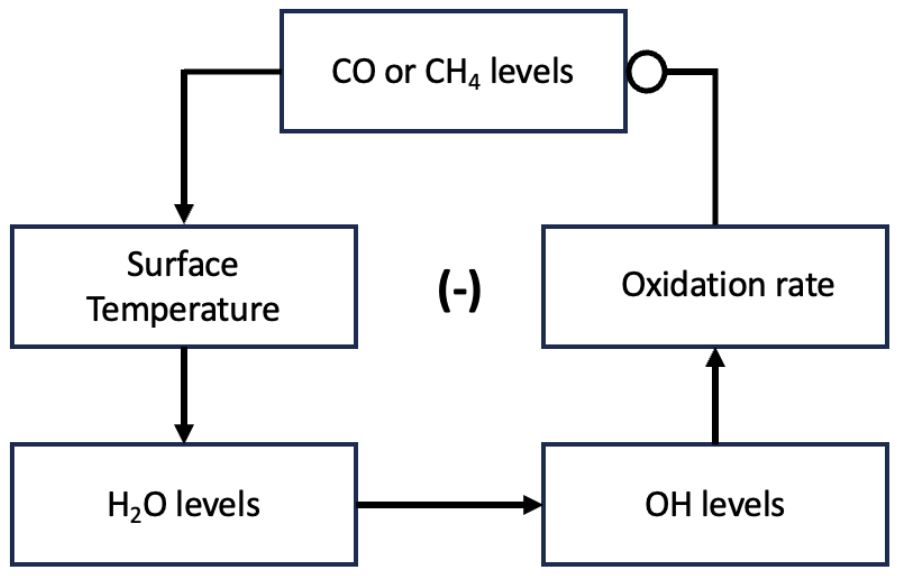}
        \put(-235,180){\scriptsize \textbf{(b)}}
        \label{fig:negative_feedback}
    \end{minipage}

    \caption{\footnotesize Qualitative climate-photochemistry feedback in the atmospheres of Earth-like planets considered in this study. A barbed arrow represents a positive coupling where an increase (decrease) in the source component yields an increase (decrease) in the target component. A circular arrow represent a negative coupling where an increase (decrease) in the source component yields a decrease (increase) in the target component.}
    \label{fig:feedback_figure}
\end{figure}

When the influence of photochemistry is coupled with the findings from the climate calculations, feedback that initiates in the atmospheres of these target terrestrial planets will cause different forms of atmospheric evolution. The feedback we are referring to is given by Figure \ref{fig:feedback_figure}. A positive feedback is shown in Figure \ref{fig:feedback_figure}a, where the addition of CO or CH\textsubscript{4} causes cooling in the atmosphere, and Figure \ref{fig:feedback_figure}b shows  a negative feedback that occurs when CO or CH\textsubscript{4} cause warming in the atmosphere.

\setlength{\abovecaptionskip}{15pt}
\setlength{\belowcaptionskip}{0pt}

First, in the fixed $p_\mathrm{CO_2} = 0.01\ \mathrm{bar}$ case (Section \ref{subsec:results:COCH4}), we expect both positive and negative feedback. A positive feedback exists in the parameter region where $p_\mathrm{CO}$ \(>\) 1 bar and $p_\mathrm{CH_4}$ \(<\) 1 bar (Figure \ref{fig:results:COCH4}a). In this region, increasing CO causes cooling because of Rayleigh scattering. The cooling then decreases water vapor levels, because the saturation vapor pressure is temperature-dependent as discussed above. Thus, there is less OH in the atmosphere and therefore a slower oxidation rate of CO \citep[e.g.,][]{kasting1990bolide}, because OH is a product of H\textsubscript{2}O photodissociation. This helps CO to build up, assuming that there is CO supply to the atmosphere from outgassing, photochemical reactions, or both (Figure \ref{fig:feedback_figure}a).

There is another positive feedback in the region of the parameter space where $p_\mathrm{CH_4} \gtrsim 10\ \mathrm{bar}$ and $p_\mathrm{CO} \lesssim 1\ \mathrm{bar}$ (Figure \ref{fig:results:COCH4}a). In this region, rising CH\textsubscript{4} levels, assuming there is also a flux from, for example, hydrothermal systems or photochemistry, would result in a temperature decrease, because of decrease in the H\textsubscript{2}O greenhouse effect (Figures \ref{fig:results:COCH4}c and \ref{fig:results:COCH4}d). The decrease of water vapor leads to less oxidation of CH\textsubscript{4} with OH, allowing further CH\textsubscript{4} buildup.

The remaining region of the parameter space, $p_\mathrm{CH_4}$ \(<\) 10 bar and $p_\mathrm{CO}$ \(<\) 1 bar, contains negative feedback. This feedback exists between both CO and CH\textsubscript{4}, and H\textsubscript{2}O. In these remaining spaces, increasing either CO or CH\textsubscript{4} will lead to the warming of the atmosphere, which subsequently raises the water vapor content (see column density figures). As was discussed previously, this would raise the oxidation rate of these gases through reactions with OH produced via H\textsubscript{2}O photodissociation, preventing further increase in CO or CH\textsubscript{4} levels (Figure \ref{fig:feedback_figure}b). This negative feedback may put upper limits on $p_\mathrm{CO}$ and $p_\mathrm{CH_4}$. 

Our results suggest that worlds with low CO\textsubscript{2} abundances are favorable for the buildup of other reducing compounds. The high $p_\mathrm{CO_2}$ case shows warming trends with increasing CO and CH$_4$ (Section \ref{subsec:results:CO2}, Figure \ref{fig:results:highCO2}a). This suggests that the parameter space for negative feedback expands compared to the low $p_\mathrm{CO_2}$ case, which may limit the buildup of other reducing compounds.

For planets orbiting M-type stars (Section \ref{subsec:results:Spec}), our results suggest dominance of negative feedback. We found that CO and CH\textsubscript{4} only act as warmers in the atmosphere (Figure \ref{fig:results:Mstar}a). Here, the greenhouse and indirect greenhouse effects of CH\textsubscript{4} and CO, respectively, would lead to increased water vapor levels.
The warming leads to increase in the H\textsubscript{s}O mixing ratio, and consequently, OH production. Then, increased oxidation rates of CO and CH\textsubscript{4} with OH may limits further buildup of these reducing species.

However, we also note that low near UV (NUV) fluxes of M-type stars support the buildup of CO and CH\textsubscript{4}. Around M-type stars, the dissociation of water vapor is slower because UV radiation is less intense in the part of the spectrum relevant to H\textsubscript{2}O dissociation ($\leq$200 nm) compared to G-type stars \citep{Tian+2014,harman2015abiotic}. Additionally, between 200 and 240 nm, HO\textsubscript{2} and H\textsubscript{2}O\textsubscript{2} are photodissociated and produce OH. With a weaker spectrum from the host star, these species are dissociated to OH less. Even in situations where carbon species cause warming and thus a negative feedback, the necessary flux of reducing species to overcome such feedback would be lower compared to that for planets orbiting G-type stars. Because of this, it is possible that these worlds can be rich in CO or CH\textsubscript{4} at lower reducing fluxes than worlds orbiting G-type stars. This suggests that CO runaway \citep{harman2015abiotic,watanabe2024relative} can happen even when CO causes warming. Furthermore, we also expect that flaring of active M-dwarf stars \citep{lloyd+2018} specifically would enhance the negative feedback because of the increased UV flux shortward of 120 nm, where H\textsubscript{2}O photodissociates and provides the OH used in oxidation of CO and CH\textsubscript{4} \citep{chang+2021}.

In the case without a carbonate-silicate cycle (fixed total carbon content in the atmosphere; Section \ref{subsec:results:pCtotal}), we suggest that CO buildup is promoted by a positive feedback, whereby cooling should decrease water vapor levels and the oxidation rate of CO, allowing for it to buildup, given a certain reducing flux from outgassing. There may exist a limit on CH\textsubscript{4} growth from positive feedback. If the ratio of CH\textsubscript{4} to CO\textsubscript{2} were to exceed unity because of disequilibrium chemistry, then it is possible for CH\textsubscript{4} to buildup in the atmosphere.

\subsection{Water Loss}\label{subsec:DiscussionWater}

\setlength{\abovecaptionskip}{15pt}
\setlength{\belowcaptionskip}{0pt}

\begin{figure}[ht]
    \centering
    \begin{minipage}[b]{0.49\textwidth}
        \centering
        \includegraphics[width=\textwidth]{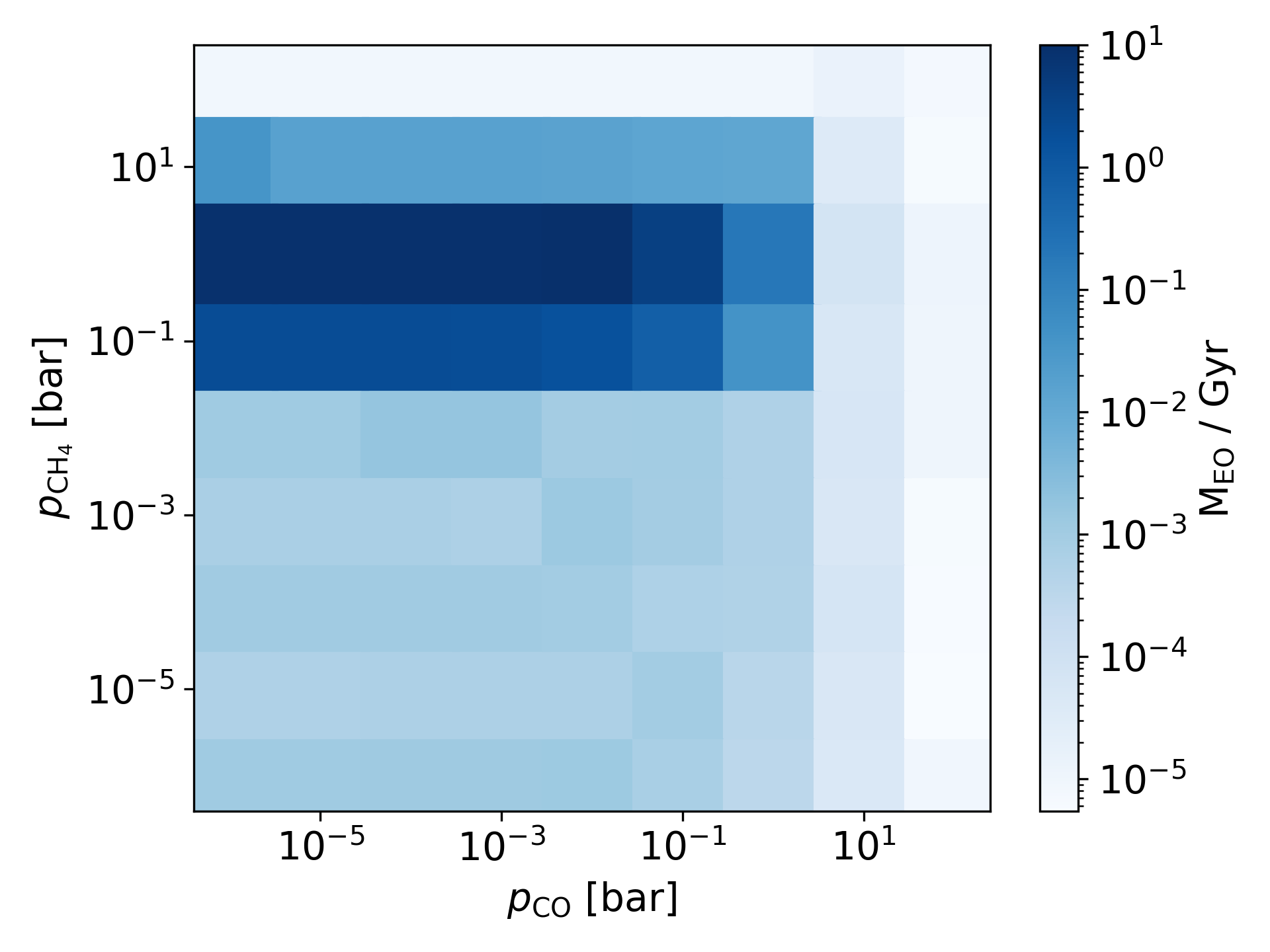}
        \put(-235,180){\scriptsize \textbf{(a)}}
        \label{fig:CO_CH4_OcLoss}
    \end{minipage}
    \hfill
    \begin{minipage}[b]{0.49\textwidth}
        \centering
        \includegraphics[width=\textwidth]{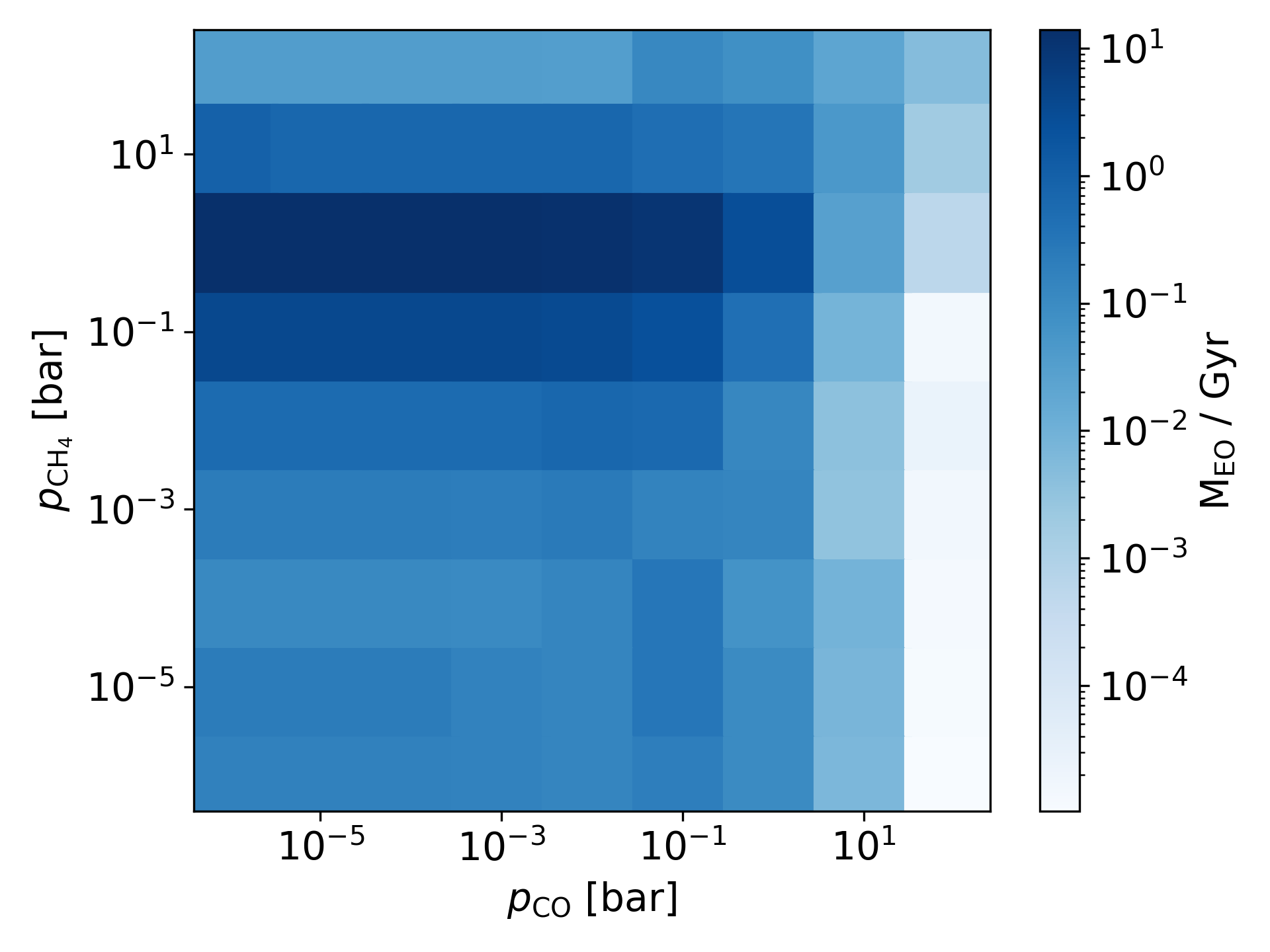}
        \put(-235,180){\scriptsize \textbf{(b)}}
        \label{fig:High_CO2_OcLoss}
    \end{minipage}
    \begin{minipage}[b]{0.49\textwidth}
        \centering
        \includegraphics[width=\textwidth]{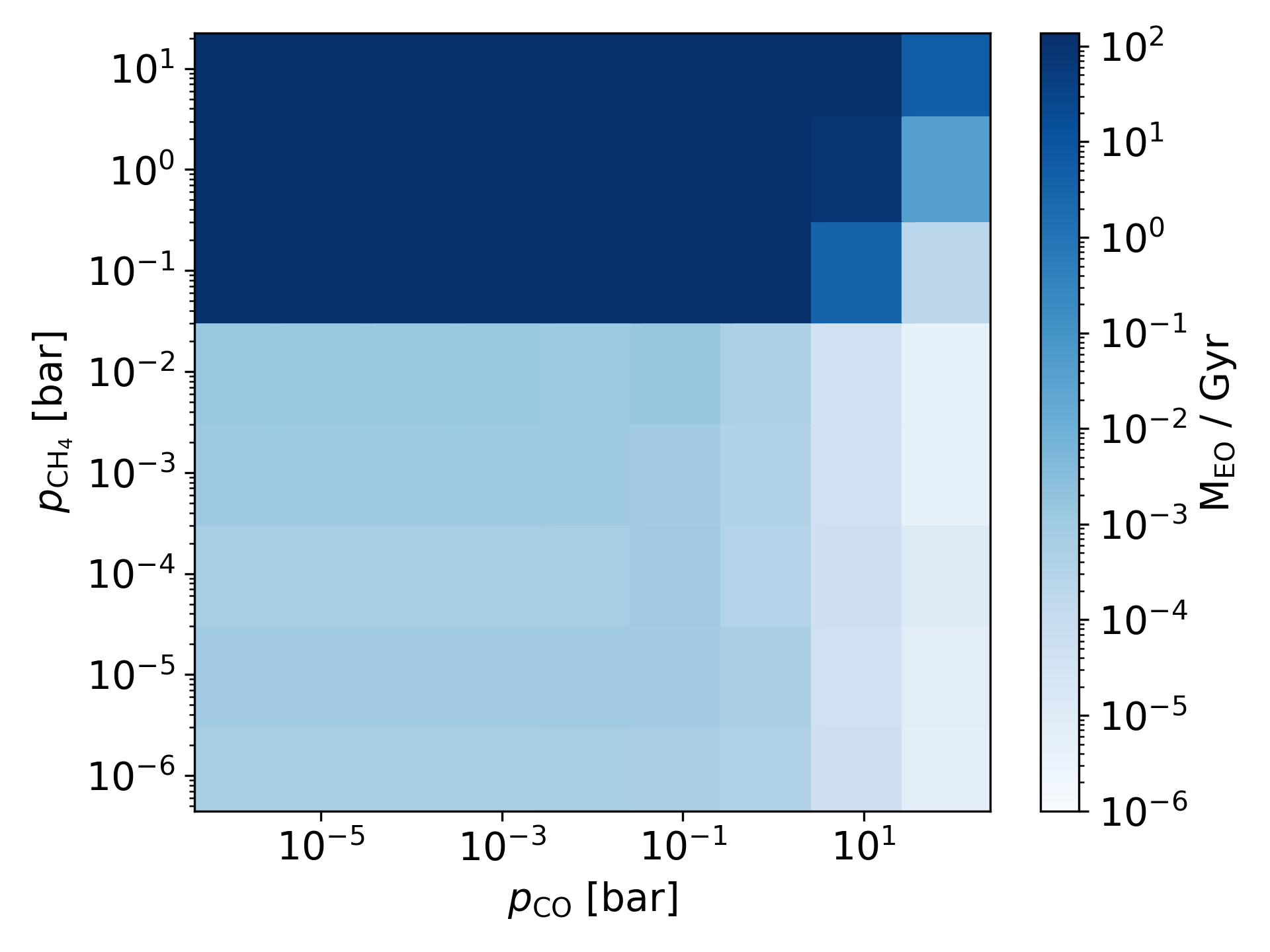}
        \put(-235,180){\scriptsize \textbf{(c)}}
        \label{fig:Mstar_OcLoss}
    \end{minipage}
    \hfill
    \begin{minipage}[b]{0.49\textwidth}
        \centering
        \includegraphics[width=\textwidth]{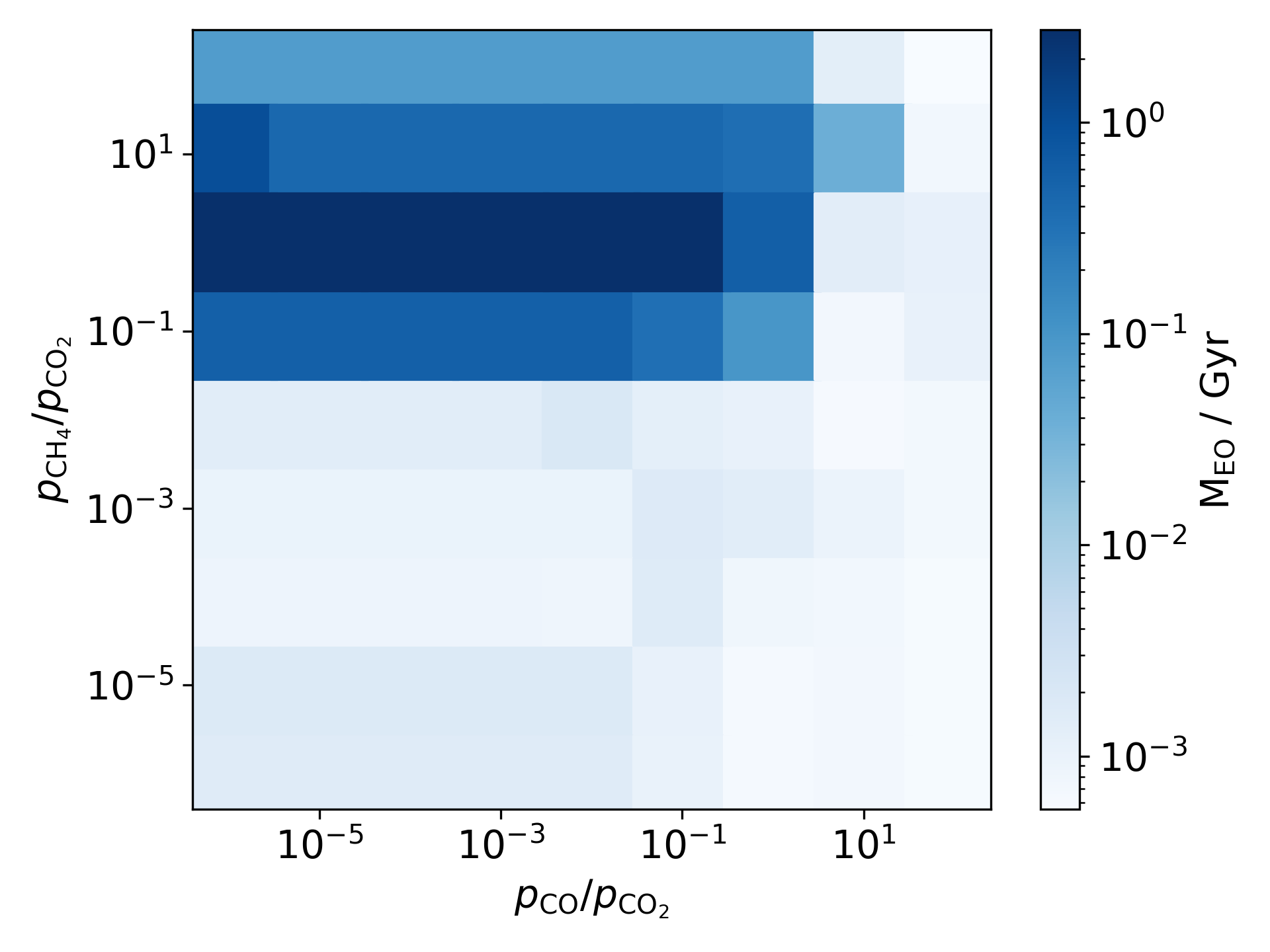}
        \put(-235,180){\scriptsize \textbf{(d)}}
        \label{fig:Fixed_OcLoss}
    \end{minipage}

    \caption{\footnotesize Dependence of water loss on the partial pressures of CO$_{2}$, CO, and CH$_{4}$; spectral type; and active carbon transfer, for an Earth-like terrestrial planet. The relationship for a Sun-like host star with 0.01 bar CO$_{2}$ is shown in (a), the relationship for a Sun-like host star with 1 bar CO$_{2}$ is shown in (b), the relationship for a GJ 876 host star with 0.01 bar CO$_{2}$ is shown in (c), and the relationship with a Sun-like host star with 0.1 bar \(\textit{p}_{\mathrm{Ctotal}}\) is shown in (d).}
    \label{fig:OcLoss_figure}
\end{figure}

In addition to the various photochemical reactions in the atmosphere that influence a planet's evolution, the loss of hydrogen from water vapor may result in the loss of water in both the atmosphere and ocean. The escape of hydrogen is thought to be regulated by two limiting factors; diffusion efficiency and the available energy \citep{catling2017atmospheric}.

The amount of water vapor at the cold trap determines the rate at which it diffuses to the upper atmosphere \citep{catling2017atmospheric}. This diffusion of hydrogen to the upper atmosphere ($\phi_l$) is calculated via the following equation; 
\begin{equation}
    \phi_l = cf_T(H),
    \label{eq:5}
\end{equation}
where $c$ is a constant calculated based on atmospheric scale height and the diffusion coefficient, given to be  
$(2.5 \times 10^{13} \,\mathrm{cm}^{-2} \,\mathrm{s}^{-1})$,  
and $f_T(H)$ is the total mixing ratio of hydrogen in all forms above the cold trap.

Then the XUV radiation from the host star determines the escape rate of diffused hydrogen ($\phi_{el}$) \citep{catling2017atmospheric}, calculated with the following equation; 
\begin{equation}
    \phi_{el} = \frac{S_{EUV}}{GMm/r},
    \label{eq:6}
\end{equation}
where $S_{EUV}$ is the incoming globally averaged extreme UV flux (EUV), $G$ is the universal gravitation constant, $M$ is the mass of the planet, $m$ is the hydrogen atom mass, and $r$ is the relevant radius for atmospheric escape.
Next, we calculate the area integrated escape rate from the Earth's surface area, divide by the mass of hydrogen in the Earth's oceans, and convert to the desired units of M\textsubscript{EO}/Gyr (where $M_\mathrm{EO}$ is the amount of water equal to the mass of hydrogen in the Earth’s ocean). From these two equations, we set the escape of hydrogen equal to whichever rate is slower. This gives us the amount of hydrogen a planet can lose relative to the Earth's oceans over the course of 1 Gyr (Figure \ref{fig:OcLoss_figure}).

The relationship between the water vapor mixing ratio and the diffusion-limited escape is linear, and the energy limit in our study is dependent only on spectral type (in reality, it also depends on the age of the star, the distance of the planet from the star, and the planet mass and radius, but all of those are fixed in our model to be consistent with a planet receiving the same flux as the modern Earth that would yield modern surface temperatures). Thus, at some point, an increase of water vapor in the atmosphere will cause the diffusion limit to exceed the energy limit, and no additional hydrogen can be lost regardless of the water vapor mixing ratio. For the case in which the Sun is the host star, the upper limit on water loss in one Earth lifetime from the energy-limited escape is 136 $M_\mathrm{EO}$, and for the M-type star case, it is 391 $M_\mathrm{EO}$ (see Figure \ref{fig:OcLoss_figure}), owing to a stronger XUV flux that allows for higher escape rates. 

In the case of the Sun as the host star (Figures \ref{fig:OcLoss_figure}a and \ref{fig:OcLoss_figure}b), increasing the methane concentration will cause water loss equivalent to $\gtrsim\ 1$ M\textsubscript{EO}/Gyr, peaking at $p_\mathrm{CH_4}$ = 1 bar. Above this value, as was discussed earlier, the amount of water vapor in the stratosphere is diminished, and thus the supply of hydrogen to the upper atmosphere is lower. A similar behavior occurs for the cases of fixed carbon content when $p_\mathrm{CH_4}$/$p_\mathrm{CO_2}$ is increased (Figure \ref{fig:OcLoss_figure}d). However, the decrease in water loss above the peak at $p_\mathrm{CH_4}$/$p_\mathrm{CO_2}$ = 1 is due to a decrease in CO\textsubscript{2} levels that lower temperatures and thus water vapor mixing ratios. In the M-type case, there is no CH\textsubscript{4} peak. Instead, increasing methane increases the amount of water that will subsequently be lost (Figure \ref{fig:OcLoss_figure}c).

In all situations, increasing the amount of CO, either through $p_\mathrm{CO}$ or $p_\mathrm{CO}$/$p_\mathrm{CO_2}$, will cause the amount of water loss to decrease. In situations where CO can cause significant warming, there is minimal risk of water loss and oxidation of the atmosphere. This suggests that terrestrial planets around M-type stars may retain surface water, potentially being habitable. We also find that high methane concentrations, roughly $p_\mathrm{CH_4}$ = 1 bar or $p_\mathrm{CH_4}$/$p_\mathrm{CO_2}$ = 1 result in high water vapor mixing ratios that ultimately cause the loss of vast amounts of water, on the order of 10 times more than the current oceans of the Earth and potentially more. This is especially true for planets orbiting M-type stars, where the upper limit from energy escape to bound loss of water is about three times higher, and where the increase of $p_\mathrm{CH_4}$ increases water vapor levels significantly. 
 
From these findings, planets with smaller CH\textsubscript{4} are favorable for water retention, unless $p_\mathrm{CO}$ \(\gtrsim\) $p_\mathrm{CH_4}$. CO rich worlds, possibly the early Earth and Mars or planets orbiting M-type stars, would still be capable of retaining significant amounts of water despite CO warming. However, certain feedback in the atmosphere may prevent these ideal situations from occurring, as was outlined in the previous section.

\subsection{Applications to Planets}\label{subsec:Discussion}

Various planets likely had or have a combination of these three carbon species and different styles of carbon cycling at some point in their history. In this subsection, we discuss the implications on atmospheric evolution of (potentially) habitable worlds: early Earth, early Mars, early Venus, and habitable exoplanets orbiting M-type stars. We note that atmospheric modeling dedicated to each planet is beyond the scope of this study; here we focus on implications which can be derived from model results for the setting of Earth-like parameters obtained in Section \ref{sec:results}.

\subsubsection{Early Earth}\label{subsubsec:Earth}

Early Earth before the Great Oxidation Event sustained a reducing atmosphere with variable amounts of CH\textsubscript{4} and, possibly, CO with a high CO\textsubscript{2} level ($p_\mathrm{CO_2} \sim $0.01--1 bar) \citep{CatlingArchean}. A reducing flux from the mantle \citep{Aulbach+2016,nicklas2019secular}, CH\textsubscript{4} production by methanogens \citep{pavlov2001uv,stueeken2020mission}, and asteroid impacts \citep{kasting1990bolide, ZahnleCreation} are thought to have contributed to sustain the reducing gases. \cite{CatlingArchean} compiled geochemical constraints \citep{Zahnle+2006,Zahnle+2019} and model predictions \citep{Claire+2006} for the CH\textsubscript{4} level and estimated $p_\mathrm{CH_4}\sim 10^{-5}$--$10^{-2}$ bar or even higher. Carbon monoxide is thought to have been a minor species, but a transient high-CO atmosphere ($p_\mathrm{CO}\sim 1\ \mathrm{bar}$) is possible after large impacts \citep{kasting1990bolide,ZahnleCreation}. 

The early Earth might have had plate tectonics from a young age, though the exact time of the onset of plate tectonics is not widely agreed upon. Studies suggest plate tectonics began sometime between 0.7 and 4.4 Ga, with evidence from ophiolites, blueschists, zircons, and various other supposedly tectonically deformed materials \citep{Harrison+2005,2005Stern,Hopkins+2008,VanHunen+2008,Foley2018}. Earlier onset of plate tectonics was also suggested from the potential accretionary complex in the 3.8-Ga-old Isua supracrustal belt \citep{Komiya+1999}.

For an early Earth with active tectonics (resulting in carbonate-silicate cycling) with a reducing flux to the atmosphere, we can apply our findings in this study to investigate atmospheric evolution. In a theoretical CH\textsubscript{4}-rich atmosphere, the transition from CH\textsubscript{4} to CO\textsubscript{2} rich would yield decreased surface temperatures and water vapor mixing ratios, as shown in Figures \ref{fig:results:COCH4} and \ref{fig:results:highCO2}. Over time, unless $p_\mathrm{CO_2}$ increases against the change in $p_\mathrm{CH_4}$, a negative feedback (Figure \ref{fig:feedback_figure}b) would self-regulate atmospheric $p_\mathrm{CH_4}$, given a sufficient outgassing flux. In the case of CO, its expected levels would likely not be large enough to incite any feedback in climate through oxidation over time. Early Earth's atmosphere may potentially (since what we estimate here is a minimum value, and thus the true value may be orders of magnitude higher for a reducing atmosphere) have been safe from significant amounts of water loss (Figures \ref{fig:OcLoss_figure}a and \ref{fig:OcLoss_figure}b), as discussed in Section \ref{subsec:DiscussionWater}.

\subsubsection{Early Mars}\label{subsubsec:Mars}

Early Mars is known to have possessed a warm climate and a hydrological cycle, at least transiently (e.g., \cite{Wordsworth2016,Ehlmann+2016}). Climate models and geochemical and geological constraints suggest $p_\mathrm{CO_2} \sim 10^{-1}$--$10^0$ bar around 3.6 to 4.5 Ga \citep{Forget+2013,wordsworth2013water,ramirez2014can,Kite+2017,kurokawa2018lower}. Because of low temperature, early Mars is prone to CO runway \citep{zahnle2008photochemical}; the CO runway might lead to $p_\mathrm{CO}$ comparable to $p_\mathrm{CO_2}$. Moreover, recent studies suggest a high background N\textsubscript{2} level ($p_\mathrm{N_2}\simeq $0.1--0.5 bar; \cite{Hu+Thomas2022}).

It is widely believed that Mars has not exhibited plate tectonics through its history. There is a lack of unambiguous evidence to suggest plate tectonics, and a stagnant lid regime is more likely \citep{Breur+2003}. 

In a stagnant lid regime, the supply of carbon species to the atmosphere of Mars would be limited. The lifetime of these species would be dictated by photochemical reactions and feedback with climate. The necessary pressure of greenhouse gasses ($p_\mathrm{GHG}$) for CO warming increases as the planet's orbital distance increases, and based on the likely composition of the ancient Martian atmosphere, CO warming would not have been possible. This puts the atmosphere in the positive feedback regime (Figure \ref{fig:feedback_figure}a) for CO (see the high $p_\mathrm{CO}$ low $p_\mathrm{CH4}$ region of Figure \ref{fig:results:COCH4}a), and would have potentially allowed for CO buildup in the atmosphere. However, the modern atmosphere is depleted in CO, suggesting that some event must have occurred to remove atmospheric CO and convert it to CO\textsubscript{2}.

\subsubsection{Early Venus}\label{subsubsec:Venus}

Different models have been proposed for the atmosphere and climate of early Venus. While some studies suggested that water never condensed on Venus \citep{Gillmann+2009,Gillman+2020,Hamano+2013,Turbet+2021}, others proposed a habitable early-Venus scenario \citep{Way+2016GeoRL..43.8376W,Way+2020}. Here we focus on the latter case, where our model results are applicable. Given poor constraints on atmospheric composition in the habitable Venus scenario, Way (2016) assumed modern-Earth-like CO\textsubscript{2} and CH\textsubscript{4} concentrations (400 ppm and 1 ppm, respectively) with the background 1 bar N\textsubscript{2}. Carbon monoxide is not usually considered, because it is not infrared active. Similar to Mars, Venus likely has always lacked plate tectonics \citep{Nimmo+1998,Lammer+2018}.

In this scenario, we expect a negative feedback to self regulate CO\textsubscript{2} and CH\textsubscript{4} levels (Figure \ref{fig:negative_feedback}), as CH\textsubscript{4} is acting as a greenhouse gas in this situation. However, at higher instellations compared to the Earth, CH\textsubscript{4} might not be capable of cooling the atmosphere, similar to the case in Figure \ref{fig:results:Mstar}a, where greenhouse processes dominate across the entire parameter space. Therefore, it is unlikely a positive feedback could operate on early Venus and allow for the buildup of CH\textsubscript{4}. We do, however, anticipate that an early Venus with abundant CO\textsubscript{2} and CH\textsubscript{4} would be susceptible to large amounts of water loss (Figures \ref{fig:OcLoss_figure}a and \ref{fig:OcLoss_figure}b).

\subsubsection{Exoplanets orbiting M-type stars}\label{subsubsec:Exoplanets}

Though Earth-sized planets orbiting habitable zones of M-type stars are primary targets for studying habitable worlds in exoplanetary systems, their atmospheric compositions are poorly constrained so far.  For instance, transmission spectra of habitable-zone Earth-sized planets orbiting TRAPPIST-1 (planets d, e, f, and g) obtained with the Hubble Space Telescope show no clear features of atmospheric gases \citep{deWit+2018}, awaiting further constraints from the James Webb Space Telescope (JWST) and future telescopes. The JWST successfully constrained atmospheric pressures for the inner planets TRAPPIST-1 b and c with secondary-eclipse observations \citep{Greene+2023,Zieba+2023}, implying potentially small volatile inventory for this system.

Constraining the presence/absence of plate tectonics on extrasolar rocky planets is more challenging. In a statistical level, it has been proposed that a correlation between stellar irradiation and $p_\mathrm{CO_2}$ predicted with active carbonate-silicate cycling can be tested with tens of Earth-like exoplanet samples \citep{Lehmer+2020,Foley2024}. Theoretical prediction for the plate tectonics on these planets is currently limited by our knowledge on how it depends on planetary parameters (such as water content and thermal history, see \cite{Wordsworth+2022} and references therein) and by uncertainty in these parameters themselves. Previous studies modeling atmospheric photochemistry of terrestrial planets orbiting M-type stars practically fixed $p_\mathrm{CO_2}$ in their parameter surveys \citep{Hu+2020,watanabe2024relative}, which may corresponds to the case with $p_\mathrm{CO_2}$ regulation with plate tectonics and an active carbonate-silicate cycle.

CO on these worlds requires lower 
$p_\mathrm{GHG}$ to act as an indirect warmer in the atmosphere (see Figures \ref{fig:results:COCH4}a and \ref{fig:results:Mstar}a). This occurs because Rayleigh scattering by CO is weaker around M-type stars (Figure \ref{fig:results:Mstar}b). This would effectively put most CO-rich worlds in the negative feedback regime, where CO levels would self-regulate themselves and be robust against buildup, unless there is a large reducing flux from the interior. The range for CH\textsubscript{4} cooling would instead occur at lower amounts of $p_\mathrm{CO_2}$, where there is a lessened greenhouse effect to be counteracted.

\subsection{Limitations}\label{subsec:discussion:limitation}

As stated in previous sections, the understanding and implementation of methane hazes are limited. We note that, in situations where $p_\mathrm{CH_4}$/$p_\mathrm{CO_2}$ \(\gtrsim\) 1, methane haze may form. On planets orbiting G-type stars, this may cause warming through the absorption of outgoing planetary radiation, or scattering of incoming radiation from the host star. However, around M-type stars, the range where these hazes would potentially absorb or scatter (between 0.4 to 1 $\mu$m) is weaker (Figure \ref{fig:methods:spectra}, and the effects of haze is greatly diminished. In such scenarios, they can be ignored \citep{harman2015abiotic,arney2017pale}. In the cases where haze would form, we may need to revisit our calculations in the future when our understanding is improved.

There are two potentially critical assumptions made for our treatment of water loss that need to be addressed. First, we assume that the atmosphere is isothermal above the tropopause. In reality, the tropopause and stratospheric temperatures will vary with height and composition \citep{wordsworth2013water}, and stratospheric temperatures may be lower than the 200 K isothermal case we have assumed. Because of the temperature dependence of water vapor on temperature, not only would water vapor levels decrease, the amount of water loss would also decrease.

Second, our discussion in Section \ref{subsec:DiscussionWater} considers only water loss as a result of mechanisms operating in the stratosphere. In reality, the feedback discussed in Section \ref{subsec:DiscussionFeedback} also occurs in the lower atmosphere. Increasing tropospheric water content in a reducing atmosphere can lead to oxidation of CO and CH\textsubscript{4}. Unless the OH produced from H\textsubscript{2}O photodissociation recombines with free hydrogen, OH will be used to oxidize the reducing species and the remaining H will escape to space. This provides a lower limit on the estimated escape of water from our modeled atmospheres. How this would be counteracted by the limitation of an isothermal stratosphere is something that may need to be revisited in the future.

Our methodology for determining the incoming stellar flux to the top of the atmosphere largely influences the findings made in our study. If the modeled planet were moved closer or farther away (to increase or decrease the stellar TOA flux, respectively), it would change the trends and implications of our study.

Finally, it is necessary to discuss the ability of our model to be applied to synchronous rotators around M-dwarf planets, specifically for the global temperature and mixing ratio distribution. The well-mixing of species is dependent on two things; the timescale of advection, and the photochemical lifetime of the given species. For an Earth-like terrestrial planet, the advection timescale of carbon species will be faster than their photochemical lifetime, as previous studies have found \citep{yates+2020,braam+2022}. This will allow the species to be homogeneously distributed globally. Numerous 3-D global climate models (GCMs) have been applied to synchronous rotators like the planets of the TRAPPIST-1 system and have investigated the variation of dayside and nightside temperatures, finding significant differences \citep{wolf2017,turbet+2018}. In the context of planetary habitability, the assumption of a global mean surface temperature is at times insufficient. \cite{lobo+2024} found that the fractional habitability of synchronous rotators around M-dwarfs varied from 16 to 79\%, suggesting that the applicability of a global mean surface temperature can vary greatly. Lastly, because the saturation vapor pressure of water is temperature dependent and thus may vary more globally than our carbon species, it may lead to varying levels of water loss on the day and night sides, respectively, which is something a 3-D GCM may be able to better model, though we anticipate the trends we find to be robust.

\section{Conclusions} \label{sec:Conclusion}

We studied the climate effects of CO\textsubscript{2}, CO, and CH\textsubscript{4} in the atmosphere of a terrestrial planet orbiting different types of host stars. We also aimed to understand the effects of CO, as recent studies suggest that CO-rich atmospheres may exist on early Mars and extrasolar terrestrial planets, especially those orbiting M-type stars. In this study, we updated a 1D atmospheric climate model of CLIMA/ATMOS to include optical and thermodynamic properties of CO. We calculated the equilibrium temperature and water vapor profiles and the surface temperature for planets with varying levels of atmospheric CO\textsubscript{2}, CO, and CH\textsubscript{4} orbiting G- and M-type stars. 

We found that, even for a CO-rich atmosphere ($p_\mathrm{CO} = 1\ \mathrm{bar}$ and $p_\mathrm{CO_2} = 10^{-3}\ \mathrm{bar}$), the impact of absorption by CO on the surface temperature is negligible. However, the absorption of stellar light by the given amount of CO increases the stratospheric temperature moderately (20--30 K) and, consequently, the stratospheric water vapor mixing ratio by an order of magnitude. Under Earth-like $p_\mathrm{N_2}$, Our parameter survey with fixed $p_\mathrm{CO_2}$ showed that increasing $p_\mathrm{CO}$ leads to surface cooling on planets orbiting Sun-like stars unless the sum of $p_\mathrm{CO_2}$ and $p_\mathrm{CH_4}$ exceeds $\sim$1 bar. This is likely caused by changing the balance between two warming mechanisms (the pressure broadening of absorption lines and increasing H$_2$O content) and cooling by Rayleigh scattering. Increasing $p_\mathrm{CH_4}$, in contrast, chiefly causes surface warming except for low $p_\mathrm{CO_2}$ ($10^{-2}\ \mathrm{bar}$) and high $p_\mathrm{CH_4}$ ($\gtrsim 10^{1}\ \mathrm{bar}$) cases, whereas organic haze formation will lead to a lower $p_\mathrm{CH_4}$ for the transition from warming to cooling. Changing the host star to a M-type star GJ 876 b leads to the dominance of warming with increasing both $p_\mathrm{CO}$ and $p_\mathrm{CH_4}$. This is likely caused by the decrease in the planetary albedo due to reduced Rayleigh scattering and absorption of stellar light by CH\textsubscript{4}. Finally, our parameter survey with fixed total carbon content shows that cooling increases with increasing $p_\mathrm{CO}/p_\mathrm{CO_2}$, and warming peaks at $p_\mathrm{CH_4}/p_\mathrm{CO_2}$ approaching unity.

The warming and cooling trends with increasing reduced species (CO and CH\textsubscript{4}) may induce negative and positive climate-photochemistry feedback, respectively. Warming (cooling) causes increase (decrease) in atmospheric water vapor content, which then decreases (increases) the oxidation rates of reduced species by OH radicals. We discussed that such climate-photochemistry feedback may have influenced the evolution of the solar and extrasolar terrestrial worlds. 

Moreover, our minimum estimate on water loss based on the obtained water vapor mixing ratio in the stratosphere showed
that atmospheres will become significantly oxidized through hydrogen loss when $p_\mathrm{CH_4}$ is approximately 1 bar, or when the methane to carbon dioxide ratio approaches unity. Around these regions, the water vapor mixing ratio in the stratosphere is at a maximum and thus is readily supplied to the upper atmosphere where it then escapes. At these peak levels we find that such planets would lose substantial amounts of water from hydrogen loss, as much as 100 times the mass of water in Earth’s oceans. Planets dominated by CO will not lose much water from their surface or atmosphere, even when CO acts as a warming gas, and planets without active transfer from the carbon cycle are also capable of retaining more hydrogen in the form of H\textsubscript{2}O. These findings help us further the understanding we have for Earth-like terrestrial planets with a wide range of possible carbon compound abundances, that we now think are both plausible and likely on Earth, Mars, and exoplanets, giving us insight into the environments they might have formed in or evolved through.

\section*{\textbf{Acknowledgements}} \label{sec:Acknowledgements}

This study was supported by JSPS Grant-in-Aid No. 20KK0080, 22H01290, 22H05150, 21H04514, and 23K22561.

\section*{\textbf{Competing Interests}} \label{sec:Competing Interests}

The authors declare no competing interests.

\bibliography{reference}{}

\begin{thebibliography}{}
\expandafter\ifx\csname natexlab\endcsname\relax\def\natexlab#1{#1}\fi
\providecommand{\url}[1]{\href{#1}{#1}}
\providecommand{\dodoi}[1]{doi:~\href{http://doi.org/#1}{\nolinkurl{#1}}}
\providecommand{\doeprint}[1]{\href{http://ascl.net/#1}{\nolinkurl{http://ascl.net/#1}}}
\providecommand{\doarXiv}[1]{\href{https://arxiv.org/abs/#1}{\nolinkurl{https://arxiv.org/abs/#1}}}

\bibitem[{{Arney} {et~al.}(2016){Arney}, {Domagal-Goldman}, {Meadows}, {Wolf}, {Schwieterman}, {Charnay}, {Claire}, {H{\'e}brard}, \& {Trainer}}]{arney2016pale}
{Arney}, G., {Domagal-Goldman}, S.~D., {Meadows}, V.~S., {et~al.} 2016, Astrobiology, 16, 873, \dodoi{10.1089/ast.2015.1422}

\bibitem[{{Arney} {et~al.}(2017){Arney}, {Meadows}, {Domagal-Goldman}, {Deming}, {Robinson}, {Tovar}, {Wolf}, \& {Schwieterman}}]{arney2017pale}
{Arney}, G.~N., {Meadows}, V.~S., {Domagal-Goldman}, S.~D., {et~al.} 2017, \apj, 836, 49, \dodoi{10.3847/1538-4357/836/1/49}

\bibitem[{{Aulbach} \& {Stagno}(2016)}]{Aulbach+2016}
{Aulbach}, S., \& {Stagno}, V. 2016, Geology, 44, 751, \dodoi{10.1130/G38070.1}

\bibitem[{{Braam} {et~al.}(2022){Braam}, {Palmer}, {Decin}, {Ridgway}, {Zamyatina}, {Mayne}, {Sergeev}, \& {Abraham}}]{braam+2022}
{Braam}, M., {Palmer}, P.~I., {Decin}, L., {et~al.} 2022, \mnras, 517, 2383, \dodoi{10.1093/mnras/stac2722}

\bibitem[{{Breuer} \& {Spohn}(2003)}]{Breur+2003}
{Breuer}, D., \& {Spohn}, T. 2003, Journal of Geophysical Research (Planets), 108, 5072, \dodoi{10.1029/2002JE001999}

\bibitem[{{Cartier} \& {Wood}(2019)}]{Cartier+Wood2019}
{Cartier}, C., \& {Wood}, B.~J. 2019, Elements, 15, 39, \dodoi{10.2138/gselements.15.1.39}

\bibitem[{{Catling} \& {Kasting}(2017)}]{catling2017atmospheric}
{Catling}, D.~C., \& {Kasting}, J.~F. 2017, {Atmospheric Evolution on Inhabited and Lifeless Worlds} (Cambridge University Press)

\bibitem[{{Catling} \& {Zahnle}(2020)}]{CatlingArchean}
{Catling}, D.~C., \& {Zahnle}, K.~J. 2020, Science Advances, 6, eaax1420, \dodoi{10.1126/sciadv.aax1420}

\bibitem[{{Chang} {et~al.}(2021){Chang}, {Yu}, {An}, {Luo}, {Quan}, {Zhang}, {Hu}, {Li}, {Yang}, {Chen}, {Che}, {Zhang}, {Wu}, {Xie}, {Ashfold}, {Yuan}, \& {Yang}}]{chang+2021}
{Chang}, Y., {Yu}, Y., {An}, F., {et~al.} 2021, Nature Communications, 12, 2476, \dodoi{10.1038/s41467-021-22824-7}

\bibitem[{{Claire} {et~al.}(2006){Claire}, {Catling}, \& {Zahnle}}]{Claire+2006}
{Claire}, M.~W., {Catling}, D.~C., \& {Zahnle}, K.~J. 2006, Geobiology, 4, 239, \dodoi{10.1111/j.1472-4669.2006.00084.x}

\bibitem[{{de Wit} {et~al.}(2018){de Wit}, {Wakeford}, {Lewis}, {Delrez}, {Gillon}, {Selsis}, {Leconte}, {Demory}, {Bolmont}, {Bourrier}, {Burgasser}, {Grimm}, {Jehin}, {Lederer}, {Owen}, {Stamenkovi{\'c}}, \& {Triaud}}]{deWit+2018}
{de Wit}, J., {Wakeford}, H.~R., {Lewis}, N.~K., {et~al.} 2018, Nature Astronomy, 2, 214, \dodoi{10.1038/s41550-017-0374-z}

\bibitem[{{Ehlmann} {et~al.}(2016){Ehlmann}, {Anderson}, {Andrews-Hanna}, {Catling}, {Christensen}, {Cohen}, {Dressing}, {Edwards}, {Elkins-Tanton}, {Farley}, {Fassett}, {Fischer}, {Fraeman}, {Golombek}, {Hamilton}, {Hayes}, {Herd}, {Horgan}, {Hu}, {Jakosky}, {Johnson}, {Kasting}, {Kerber}, {Kinch}, {Kite}, {Knutson}, {Lunine}, {Mahaffy}, {Mangold}, {McCubbin}, {Mustard}, {Niles}, {Quantin-Nataf}, {Rice}, {Stack}, {Stevenson}, {Stewart}, {Toplis}, {Usui}, {Weiss}, {Werner}, {Wordsworth}, {Wray}, {Yingst}, {Yung}, \& {Zahnle}}]{Ehlmann+2016}
{Ehlmann}, B.~L., {Anderson}, F.~S., {Andrews-Hanna}, J., {et~al.} 2016, Journal of Geophysical Research (Planets), 121, 1927, \dodoi{10.1002/2016JE005134}

\bibitem[{{Foley}(2015)}]{foley2015role}
{Foley}, B.~J. 2015, \apj, 812, 36, \dodoi{10.1088/0004-637X/812/1/36}

\bibitem[{{Foley}(2018)}]{Foley2018}
---. 2018, Philosophical Transactions of the Royal Society of London Series A, 376, 20170409, \dodoi{10.1098/rsta.2017.0409}

\bibitem[{Foley(2024)}]{Foley2024}
Foley, B.~J. 2024, Reviews in Mineralogy and Geochemistry, 90, 559, \dodoi{10.2138/rmg.2024.90.15}

\bibitem[{{Foley} \& {Smye}(2018)}]{foley+2018}
{Foley}, B.~J., \& {Smye}, A.~J. 2018, Astrobiology, 18, 873, \dodoi{10.1089/ast.2017.1695}

\bibitem[{{Forget} {et~al.}(2013){Forget}, {Wordsworth}, {Millour}, {Madeleine}, {Kerber}, {Leconte}, {Marcq}, \& {Haberle}}]{Forget+2013}
{Forget}, F., {Wordsworth}, R., {Millour}, E., {et~al.} 2013, \icarus, 222, 81, \dodoi{10.1016/j.icarus.2012.10.019}

\bibitem[{{Gillmann} {et~al.}(2009){Gillmann}, {Chassefi{\`e}re}, \& {Lognonn{\'e}}}]{Gillmann+2009}
{Gillmann}, C., {Chassefi{\`e}re}, E., \& {Lognonn{\'e}}, P. 2009, Earth and Planetary Science Letters, 286, 503, \dodoi{10.1016/j.epsl.2009.07.016}

\bibitem[{{Gillmann} {et~al.}(2020){Gillmann}, {Golabek}, {Raymond}, {Sch{\"o}nb{\"a}chler}, {Tackley}, {Dehant}, \& {Debaille}}]{Gillman+2020}
{Gillmann}, C., {Golabek}, G.~J., {Raymond}, S.~N., {et~al.} 2020, Nature Geoscience, 13, 265, \dodoi{10.1038/s41561-020-0561-x}

\bibitem[{{Goessling} \& {Bathiany}(2016)}]{Goessling+Bathiany2016ESD.....7..697G}
{Goessling}, H.~F., \& {Bathiany}, S. 2016, Earth System Dynamics, 7, 697, \dodoi{10.5194/esd-7-697-2016}

\bibitem[{{Goldblatt} {et~al.}(2009){Goldblatt}, {Claire}, {Lenton}, {Matthews}, {Watson}, \& {Zahnle}}]{Goldblatt+2009}
{Goldblatt}, C., {Claire}, M.~W., {Lenton}, T.~M., {et~al.} 2009, Nature Geoscience, 2, 891, \dodoi{10.1038/ngeo692}

\bibitem[{{Goldblatt} {et~al.}(2013){Goldblatt}, {Robinson}, {Zahnle}, \& {Crisp}}]{Goldblatt+2013NatGe...6..661G}
{Goldblatt}, C., {Robinson}, T.~D., {Zahnle}, K.~J., \& {Crisp}, D. 2013, Nature Geoscience, 6, 661, \dodoi{10.1038/ngeo1892}

\bibitem[{{Gordon} {et~al.}(2017){Gordon}, {Rothman}, {Hill}, {Kochanov}, {Tan}, {Bernath}, {Birk}, {Boudon}, {Campargue}, {Chance}, {Drouin}, {Flaud}, {Gamache}, {Hodges}, {Jacquemart}, {Perevalov}, {Perrin}, {Shine}, {Smith}, {Tennyson}, {Toon}, {Tran}, {Tyuterev}, {Barbe}, {Cs{\'a}sz{\'a}r}, {Devi}, {Furtenbacher}, {Harrison}, {Hartmann}, {Jolly}, {Johnson}, {Karman}, {Kleiner}, {Kyuberis}, {Loos}, {Lyulin}, {Massie}, {Mikhailenko}, {Moazzen-Ahmadi}, {M{\"u}ller}, {Naumenko}, {Nikitin}, {Polyansky}, {Rey}, {Rotger}, {Sharpe}, {Sung}, {Starikova}, {Tashkun}, {Auwera}, {Wagner}, {Wilzewski}, {Wcis{\l}o}, {Yu}, \& {Zak}}]{gordon2017hitran2016}
{Gordon}, I.~E., {Rothman}, L.~S., {Hill}, C., {et~al.} 2017, \jqsrt, 203, 3, \dodoi{10.1016/j.jqsrt.2017.06.038}

\bibitem[{{Greene} {et~al.}(2023){Greene}, {Bell}, {Ducrot}, {Dyrek}, {Lagage}, \& {Fortney}}]{Greene+2023}
{Greene}, T.~P., {Bell}, T.~J., {Ducrot}, E., {et~al.} 2023, \nat, 618, 39, \dodoi{10.1038/s41586-023-05951-7}

\bibitem[{{Hamano} {et~al.}(2013){Hamano}, {Abe}, \& {Genda}}]{Hamano+2013}
{Hamano}, K., {Abe}, Y., \& {Genda}, H. 2013, \nat, 497, 607, \dodoi{10.1038/nature12163}

\bibitem[{{Haqq-Misra} {et~al.}(2008){Haqq-Misra}, {Domagal-Goldman}, {Kasting}, \& {Kasting}}]{haqq2008revised}
{Haqq-Misra}, J.~D., {Domagal-Goldman}, S.~D., {Kasting}, P.~J., \& {Kasting}, J.~F. 2008, Astrobiology, 8, 1127, \dodoi{10.1089/ast.2007.0197}

\bibitem[{{Harman} {et~al.}(2015){Harman}, {Schwieterman}, {Schottelkotte}, \& {Kasting}}]{harman2015abiotic}
{Harman}, C.~E., {Schwieterman}, E.~W., {Schottelkotte}, J.~C., \& {Kasting}, J.~F. 2015, \apj, 812, 137, \dodoi{10.1088/0004-637X/812/2/137}

\bibitem[{{Harrison} {et~al.}(2005){Harrison}, {Blichert-Toft}, {M{\"u}ller}, {Albarede}, {Holden}, \& {Mojzsis}}]{Harrison+2005}
{Harrison}, T.~M., {Blichert-Toft}, J., {M{\"u}ller}, W., {et~al.} 2005, Science, 310, 1947, \dodoi{10.1126/science.1117926}

\bibitem[{{Hopkins} {et~al.}(2008){Hopkins}, {Harrison}, \& {Manning}}]{Hopkins+2008}
{Hopkins}, M., {Harrison}, T.~M., \& {Manning}, C.~E. 2008, \nat, 456, 493, \dodoi{10.1038/nature07465}

\bibitem[{{Hu} {et~al.}(2020){Hu}, {Peterson}, \& {Wolf}}]{Hu+2020}
{Hu}, R., {Peterson}, L., \& {Wolf}, E.~T. 2020, \apj, 888, 122, \dodoi{10.3847/1538-4357/ab5f07}

\bibitem[{{Hu} \& {Thomas}(2022)}]{Hu+Thomas2022}
{Hu}, R., \& {Thomas}, T.~B. 2022, Nature Geoscience, 15, 106, \dodoi{10.1038/s41561-021-00886-y}

\bibitem[{{Hunten}(1973)}]{Hunten1973JAtS...30.1481H}
{Hunten}, D.~M. 1973, Journal of the Atmospheric Sciences, 30, 1481, \dodoi{10.1175/1520-0469(1973)030<1481:TEOLGF>2.0.CO;2}

\bibitem[{{Ingersoll}(1969)}]{Ingersoll1969}
{Ingersoll}, A.~P. 1969, Journal of the Atmospheric Sciences, 26, 1191, \dodoi{10.1175/1520-0469(1969)026<1191:TRGAHO>2.0.CO;2}

\bibitem[{{Kadoya} {et~al.}(2020){Kadoya}, {Catling}, {Nicklas}, {Puchtel}, \& {Anbar}}]{Kadoya+2020}
{Kadoya}, S., {Catling}, D.~C., {Nicklas}, R.~W., {Puchtel}, I.~S., \& {Anbar}, A.~D. 2020, Nature Communications, 11, 2774, \dodoi{10.1038/s41467-020-16493-1}

\bibitem[{{Kasting}(1988)}]{kasting1988runaway}
{Kasting}, J.~F. 1988, \icarus, 74, 472, \dodoi{10.1016/0019-1035(88)90116-9}

\bibitem[{{Kasting}(1990)}]{kasting1990bolide}
---. 1990, Origins of Life and Evolution of the Biosphere, 20, 199, \dodoi{10.1007/BF01808105}

\bibitem[{{Kasting}(1993)}]{kasting1993earth}
---. 1993, Science, 259, 920, \dodoi{10.1126/science.259.5097.92010.1126/science.11536547}

\bibitem[{{Kasting} \& {Ackerman}(1986)}]{kasting1986climatic}
{Kasting}, J.~F., \& {Ackerman}, T.~P. 1986, Science, 234, 1383, \dodoi{10.1126/science.11539665}

\bibitem[{{Kasting} {et~al.}(1984){Kasting}, {Pollack}, \& {Ackerman}}]{kasting1984response}
{Kasting}, J.~F., {Pollack}, J.~B., \& {Ackerman}, T.~P. 1984, \icarus, 57, 335, \dodoi{10.1016/0019-1035(84)90122-2}

\bibitem[{{Kasting} {et~al.}(1993){Kasting}, {Whitmire}, \& {Reynolds}}]{Kasting+1993Icar..101..108K}
{Kasting}, J.~F., {Whitmire}, D.~P., \& {Reynolds}, R.~T. 1993, \icarus, 101, 108, \dodoi{10.1006/icar.1993.1010}

\bibitem[{{Kato} {et~al.}(1999){Kato}, {Ackerman}, {Mather}, \& {Clothiaux}}]{kato1999k}
{Kato}, S., {Ackerman}, T.~P., {Mather}, J.~H., \& {Clothiaux}, E.~E. 1999, \jqsrt, 62, 109, \dodoi{10.1016/S0022-4073(98)00075-2}

\bibitem[{{Kite} {et~al.}(2017){Kite}, {Gao}, {Goldblatt}, {Mischna}, {Mayer}, \& {Yung}}]{Kite+2017}
{Kite}, E.~S., {Gao}, P., {Goldblatt}, C., {et~al.} 2017, Nature Geoscience, 10, 737, \dodoi{10.1038/ngeo3033}

\bibitem[{{Komiya} {et~al.}(1999){Komiya}, {Maruyama}, {Masuda}, {Nohda}, {Hayashi}, \& {Okamoto}}]{Komiya+1999}
{Komiya}, T., {Maruyama}, S., {Masuda}, T., {et~al.} 1999, Journal of Geology, 107, 515, \dodoi{10.1086/314371}

\bibitem[{{Kopparapu} {et~al.}(2013){Kopparapu}, {Ramirez}, {Kasting}, {Eymet}, {Robinson}, {Mahadevan}, {Terrien}, {Domagal-Goldman}, {Meadows}, \& {Deshpande}}]{Kopparapu+2013ApJ...765..131K}
{Kopparapu}, R.~K., {Ramirez}, R., {Kasting}, J.~F., {et~al.} 2013, \apj, 765, 131, \dodoi{10.1088/0004-637X/765/2/131}

\bibitem[{{Krissansen-Totton} {et~al.}(2018){Krissansen-Totton}, {Arney}, \& {Catling}}]{Krissansen-Totton+2018PNAS..115.4105}
{Krissansen-Totton}, J., {Arney}, G.~N., \& {Catling}, D.~C. 2018, Proceedings of the National Academy of Science, 115, 4105, \dodoi{10.1073/pnas.1721296115}

\bibitem[{{Kurokawa} {et~al.}(2018){Kurokawa}, {Kurosawa}, \& {Usui}}]{kurokawa2018lower}
{Kurokawa}, H., {Kurosawa}, K., \& {Usui}, T. 2018, \icarus, 299, 443, \dodoi{10.1016/j.icarus.2017.08.020}

\bibitem[{{Lammer} {et~al.}(2018){Lammer}, {Zerkle}, {Gebauer}, {Tosi}, {Noack}, {Scherf}, {Pilat-Lohinger}, {G{\"u}del}, {Grenfell}, {Godolt}, \& {Nikolaou}}]{Lammer+2018}
{Lammer}, H., {Zerkle}, A.~L., {Gebauer}, S., {et~al.} 2018, \aapr, 26, 2, \dodoi{10.1007/s00159-018-0108-y}

\bibitem[{{Lehmer} {et~al.}(2020){Lehmer}, {Catling}, \& {Krissansen-Totton}}]{Lehmer+2020}
{Lehmer}, O.~R., {Catling}, D.~C., \& {Krissansen-Totton}, J. 2020, Nature Communications, 11, 6153, \dodoi{10.1038/s41467-020-19896-2}

\bibitem[{{Lobo} \& {Shields}(2024)}]{lobo+2024}
{Lobo}, A.~H., \& {Shields}, A.~L. 2024, \apj, 972, 71, \dodoi{10.3847/1538-4357/ad58bb}

\bibitem[{{Loyd} {et~al.}(2018){Loyd}, {France}, {Youngblood}, {Schneider}, {Brown}, {Hu}, {Segura}, {Linsky}, {Redfield}, {Tian}, {Rugheimer}, {Miguel}, \& {Froning}}]{lloyd+2018}
{Loyd}, R.~O.~P., {France}, K., {Youngblood}, A., {et~al.} 2018, \apj, 867, 71, \dodoi{10.3847/1538-4357/aae2bd}

\bibitem[{{Manabe} \& {Wetherald}(1967{\natexlab{a}})}]{Manabe+Wetherald1967}
{Manabe}, S., \& {Wetherald}, R.~T. 1967{\natexlab{a}}, Journal of the Atmospheric Sciences, 24, 241, \dodoi{10.1175/1520-0469(1967)024<0241:TEOTAW>2.0.CO;2}

\bibitem[{{Manabe} \& {Wetherald}(1967{\natexlab{b}})}]{manabe1967thermal}
---. 1967{\natexlab{b}}, Journal of the Atmospheric Sciences, 24, 241, \dodoi{10.1175/1520-0469(1967)024<0241:TEOTAW>2.0.CO;2}

\bibitem[{{McElroy} \& {Donahue}(1972)}]{McElroy+Donahue1972Sci...177..986M}
{McElroy}, M.~B., \& {Donahue}, T.~M. 1972, Science, 177, 986, \dodoi{10.1126/science.177.4053.986}

\bibitem[{{Nicklas} {et~al.}(2019){Nicklas}, {Puchtel}, {Ash}, {Piccoli}, {Hanski}, {Nisbet}, {Waterton}, {Pearson}, \& {Anbar}}]{nicklas2019secular}
{Nicklas}, R.~W., {Puchtel}, I.~S., {Ash}, R.~D., {et~al.} 2019, \gca, 250, 49, \dodoi{10.1016/j.gca.2019.01.037}

\bibitem[{{Nimmo} \& {McKenzie}(1998)}]{Nimmo+1998}
{Nimmo}, F., \& {McKenzie}, D. 1998, Annual Review of Earth and Planetary Sciences, 26, 23, \dodoi{10.1146/annurev.earth.26.1.23}

\bibitem[{{Pavlov} {et~al.}(2001){Pavlov}, {Brown}, \& {Kasting}}]{pavlov2001uv}
{Pavlov}, A.~A., {Brown}, L.~L., \& {Kasting}, J.~F. 2001, \jgr, 106, 23267, \dodoi{10.1029/2000JE001448}

\bibitem[{{Pavlov} {et~al.}(2000){Pavlov}, {Kasting}, {Brown}, {Rages}, \& {Freedman}}]{pavlov2000greenhouse}
{Pavlov}, A.~A., {Kasting}, J.~F., {Brown}, L.~L., {Rages}, K.~A., \& {Freedman}, R. 2000, \jgr, 105, 11981, \dodoi{10.1029/1999JE001134}

\bibitem[{{Pierrehumbert}(2010)}]{Pierrehumbert2010ppc..book.....P}
{Pierrehumbert}, R.~T. 2010, {Principles of Planetary Climate}

\bibitem[{{Ramirez} \& {Kaltenegger}(2018)}]{Ramirez+2018}
{Ramirez}, R.~M., \& {Kaltenegger}, L. 2018, \apj, 858, 72, \dodoi{10.3847/1538-4357/aab8fa}

\bibitem[{{Ramirez} {et~al.}(2014){Ramirez}, {Kopparapu}, {Lindner}, \& {Kasting}}]{ramirez2014can}
{Ramirez}, R.~M., {Kopparapu}, R.~K., {Lindner}, V., \& {Kasting}, J.~F. 2014, Astrobiology, 14, 714, \dodoi{10.1089/ast.2014.1153}

\bibitem[{{Rothman} {et~al.}(2009){Rothman}, {Gordon}, {Barbe}, {Benner}, {Bernath}, {Birk}, {Boudon}, {Brown}, {Campargue}, {Champion}, {Chance}, {Coudert}, {Dana}, {Devi}, {Fally}, {Flaud}, {Gamache}, {Goldman}, {Jacquemart}, {Kleiner}, {Lacome}, {Lafferty}, {Mandin}, {Massie}, {Mikhailenko}, {Miller}, {Moazzen-Ahmadi}, {Naumenko}, {Nikitin}, {Orphal}, {Perevalov}, {Perrin}, {Predoi-Cross}, {Rinsland}, {Rotger}, {{\v{S}}ime{\v{c}}kov{\'a}}, {Smith}, {Sung}, {Tashkun}, {Tennyson}, {Toth}, {Vandaele}, \& {Vander Auwera}}]{rothman2009hitran}
{Rothman}, L.~S., {Gordon}, I.~E., {Barbe}, A., {et~al.} 2009, \jqsrt, 110, 533, \dodoi{10.1016/j.jqsrt.2009.02.013}

\bibitem[{{Stern}(2005)}]{2005Stern}
{Stern}, R.~J. 2005, Geology, 33, 557, \dodoi{10.1130/G21365.1}

\bibitem[{{St{\"u}eken} {et~al.}(2020){St{\"u}eken}, {Som}, {Claire}, {Rugheimer}, {Scherf}, {Spro{\ss}}, {Tosi}, {Ueno}, \& {Lammer}}]{stueeken2020mission}
{St{\"u}eken}, E.~E., {Som}, S.~M., {Claire}, M., {et~al.} 2020, \ssr, 216, 31, \dodoi{10.1007/s11214-020-00652-3}

\bibitem[{{Tian} {et~al.}(2014){Tian}, {France}, {Linsky}, {Mauas}, \& {Vieytes}}]{Tian+2014}
{Tian}, F., {France}, K., {Linsky}, J.~L., {Mauas}, P. J.~D., \& {Vieytes}, M.~C. 2014, Earth and Planetary Science Letters, 385, 22, \dodoi{10.1016/j.epsl.2013.10.024}

\bibitem[{{Toon} {et~al.}(1989){Toon}, {McKay}, {Ackerman}, \& {Santhanam}}]{toon1989rapid}
{Toon}, O.~B., {McKay}, C.~P., {Ackerman}, T.~P., \& {Santhanam}, K. 1989, \jgr, 94, 16287, \dodoi{10.1029/JD094iD13p16287}

\bibitem[{{Turbet} {et~al.}(2021){Turbet}, {Bolmont}, {Chaverot}, {Ehrenreich}, {Leconte}, \& {Marcq}}]{Turbet+2021}
{Turbet}, M., {Bolmont}, E., {Chaverot}, G., {et~al.} 2021, \nat, 598, 276, \dodoi{10.1038/s41586-021-03873-w}

\bibitem[{{Turbet} {et~al.}(2018){Turbet}, {Bolmont}, {Leconte}, {Forget}, {Selsis}, {Tobie}, {Caldas}, {Naar}, \& {Gillon}}]{turbet+2018}
{Turbet}, M., {Bolmont}, E., {Leconte}, J., {et~al.} 2018, \aap, 612, A86, \dodoi{10.1051/0004-6361/201731620}

\bibitem[{{Ueno} {et~al.}(2024){Ueno}, {Schmidt}, {Johnson}, {Zang}, {Gilbert}, {Kurokawa}, {Usui}, \& {Aoki}}]{Ueno+2024}
{Ueno}, Y., {Schmidt}, J.~A., {Johnson}, M.~S., {et~al.} 2024, Nature Geoscience, 17, 503, \dodoi{10.1038/s41561-024-01443-z}

\bibitem[{{van Hunen} \& {van den Berg}(2008)}]{VanHunen+2008}
{van Hunen}, J., \& {van den Berg}, A.~P. 2008, Lithos, 103, 217, \dodoi{10.1016/j.lithos.2007.09.016}

\bibitem[{{Vardavas} \& {Carver}(1984)}]{vardavas1984solar}
{Vardavas}, I.~M., \& {Carver}, J.~H. 1984, \planss, 32, 1307, \dodoi{10.1016/0032-0633(84)90074-6}

\bibitem[{{Walker} {et~al.}(1981){Walker}, {Hays}, \& {Kasting}}]{walker1981negative}
{Walker}, J.~C.~G., {Hays}, P.~B., \& {Kasting}, J.~F. 1981, \jgr, 86, 9776, \dodoi{10.1029/JC086iC10p09776}

\bibitem[{{Watanabe} \& {Ozaki}(2024)}]{watanabe2024relative}
{Watanabe}, Y., \& {Ozaki}, K. 2024, \apj, 961, 1, \dodoi{10.3847/1538-4357/ad10a2}

\bibitem[{{Watson} {et~al.}(1981){Watson}, {Donahue}, \& {Walker}}]{Watson+1981Icar...48..150W}
{Watson}, A.~J., {Donahue}, T.~M., \& {Walker}, J.~C.~G. 1981, \icarus, 48, 150, \dodoi{10.1016/0019-1035(81)90101-9}

\bibitem[{{Way} \& {Del Genio}(2020)}]{Way+2020}
{Way}, M.~J., \& {Del Genio}, A.~D. 2020, Journal of Geophysical Research (Planets), 125, e06276, \dodoi{10.1029/2019JE00627610.1002/essoar.10501118.3}

\bibitem[{{Way} {et~al.}(2016){Way}, {Del Genio}, {Kiang}, {Sohl}, {Grinspoon}, {Aleinov}, {Kelley}, \& {Clune}}]{Way+2016GeoRL..43.8376W}
{Way}, M.~J., {Del Genio}, A.~D., {Kiang}, N.~Y., {et~al.} 2016, \grl, 43, 8376, \dodoi{10.1002/2016GL069790}

\bibitem[{{Way} {et~al.}(2017){Way}, {Aleinov}, {Amundsen}, {Chandler}, {Clune}, {Del Genio}, {Fujii}, {Kelley}, {Kiang}, {Sohl}, \& {Tsigaridis}}]{way+2017}
{Way}, M.~J., {Aleinov}, I., {Amundsen}, D.~S., {et~al.} 2017, \apjs, 231, 12, \dodoi{10.3847/1538-4365/aa7a06}

\bibitem[{{Wolf}(2017)}]{wolf2017}
{Wolf}, E.~T. 2017, \apjl, 839, L1, \dodoi{10.3847/2041-8213/aa693a}

\bibitem[{{Wordsworth} \& {Kreidberg}(2022)}]{Wordsworth+2022}
{Wordsworth}, R., \& {Kreidberg}, L. 2022, \araa, 60, 159, \dodoi{10.1146/annurev-astro-052920-125632}

\bibitem[{{Wordsworth}(2016)}]{Wordsworth2016}
{Wordsworth}, R.~D. 2016, Annual Review of Earth and Planetary Sciences, 44, 381, \dodoi{10.1146/annurev-earth-060115-012355}

\bibitem[{{Wordsworth} \& {Pierrehumbert}(2013)}]{wordsworth2013water}
{Wordsworth}, R.~D., \& {Pierrehumbert}, R.~T. 2013, \apj, 778, 154, \dodoi{10.1088/0004-637X/778/2/154}

\bibitem[{{Yates} {et~al.}(2020){Yates}, {Palmer}, {Manners}, {Boutle}, {Kohary}, {Mayne}, \& {Abraham}}]{yates+2020}
{Yates}, J.~S., {Palmer}, P.~I., {Manners}, J., {et~al.} 2020, \mnras, 492, 1691, \dodoi{10.1093/mnras/stz3520}

\bibitem[{{Zahnle} {et~al.}(2006){Zahnle}, {Claire}, \& {Catling}}]{Zahnle+2006}
{Zahnle}, K., {Claire}, M., \& {Catling}, D. 2006, Geobiology, 4, 271, \dodoi{10.1111/j.1472-4669.2006.00085.x}

\bibitem[{{Zahnle} {et~al.}(2008){Zahnle}, {Haberle}, {Catling}, \& {Kasting}}]{zahnle2008photochemical}
{Zahnle}, K., {Haberle}, R.~M., {Catling}, D.~C., \& {Kasting}, J.~F. 2008, Journal of Geophysical Research (Planets), 113, E11004, \dodoi{10.1029/2008JE003160}

\bibitem[{{Zahnle} {et~al.}(2019){Zahnle}, {Gacesa}, \& {Catling}}]{Zahnle+2019}
{Zahnle}, K.~J., {Gacesa}, M., \& {Catling}, D.~C. 2019, \gca, 244, 56, \dodoi{10.1016/j.gca.2018.09.017}

\bibitem[{{Zahnle} {et~al.}(2020){Zahnle}, {Lupu}, {Catling}, \& {Wogan}}]{ZahnleCreation}
{Zahnle}, K.~J., {Lupu}, R., {Catling}, D.~C., \& {Wogan}, N. 2020, \psj, 1, 11, \dodoi{10.3847/PSJ/ab7e2c}

\bibitem[{{Zhang} {et~al.}(2024){Zhang}, {Stagno}, {Zhang}, {Chen}, {Liu}, {Li}, \& {Sun}}]{Zhang+2024}
{Zhang}, F., {Stagno}, V., {Zhang}, L., {et~al.} 2024, Nature Communications, 15, 6521, \dodoi{10.1038/s41467-024-50778-z}

\bibitem[{{Zieba} {et~al.}(2023){Zieba}, {Kreidberg}, {Ducrot}, {Gillon}, {Morley}, {Schaefer}, {Tamburo}, {Koll}, {Lyu}, {Acu{\~n}a}, {Agol}, {Iyer}, {Hu}, {Lincowski}, {Meadows}, {Selsis}, {Bolmont}, {Mandell}, \& {Suissa}}]{Zieba+2023}
{Zieba}, S., {Kreidberg}, L., {Ducrot}, E., {et~al.} 2023, \nat, 620, 746, \dodoi{10.1038/s41586-023-06232-z}

\end{thebibliography}
\bibliographystyle{aasjournal}

\end{document}